\documentclass[a4paper,12pt, oneside]{article}
\usepackage{amsmath}
\usepackage{amssymb, latexsym}
\usepackage{amsfonts}
\usepackage{makeidx}
\usepackage{ascmac}
\usepackage{amssymb}

\usepackage{color}     

\date{}

\newtheorem{lem}{Lemma}[section]
\newtheorem{thm}{Theorem}[section]
\newtheorem{prop}{Proposition}[section]

\newcommand{\dbar}{d\!\!\!{\lower-0.6ex\hbox{$-$}}\!}
\newcommand{\dslash}{d\!\!\!{\lower-0.6ex\hbox{$-$}}}

\newcommand{\h}{\hbar}

\newcommand{\btt}{\lower-0.2ex\hbox{${\scriptscriptstyle{\bullet}}$}}
\newcommand{\ctt}{\lower-0.2ex\hbox{${\scriptscriptstyle{\circ}}$}}
\newcommand{\dtt}{\lower-0.2ex\hbox{${\scriptscriptstyle{\diamond}}$}}
\newcommand{\odt}{\lower-0.4ex\hbox{${\scriptscriptstyle{\odot}}$}}

\newcommand{\Cal}{\mathcal}

\newcommand{\rd}{\partial}

\newcommand{\dS}{{\mathcal S}} 
\begin{document}

\pagestyle{plain}

\title{Deformation Expression for Elements of Algebras (VI)\\
--Vacuum representation of Heisenberg algebra--}

\author{
     Hideki Omori\thanks{ Department of Mathematics,
             Faculty of Sciences and Technology,
        Tokyo University of Science, 2641, Noda, Chiba, 278-8510, Japan,
         email: omori@ma.noda.tus.ac.jp}
        \\Tokyo University of Science
\and  Yoshiaki Maeda\thanks{Department of Mathematics,
                Faculty of Science and Technology,
                Keio University, 3-14-1, Hiyoshi, Yokohama,223-8522, Japan,
                email: maeda@math.keio.ac.jp}
          \\Keio University
\and  Naoya Miyazaki\thanks{ Department of Mathematics, Faculty of
Economics, Keio University,  4-1-1, Hiyoshi, Yokohama, 223-8521, Japan,
        email: miyazaki@hc.cc.keio.ac.jp}
        \\Keio University
\and  Akira Yoshioka \thanks{ Department of Mathematics,
          Faculty of Science, Tokyo University of Science,
         1-3, Kagurazaka, Tokyo, 102-8601, Japan,
         email: yoshioka@rs.kagu.tus.ac.jp}
           \\Tokyo University of Science
     }

\maketitle

\tableofcontents

\pagestyle{plain}



\setcounter{equation}{0}

\bigskip 




\vspace{1.5cm}

\section{Introduction}\label{prel01}

The {\bf Weyl algebra}   
$(W_{2m}[\h]; *)$ is the algebra generated by 
${\pmb u}{=}({u}_1,\cdots,{u}_m,{v}_1,\cdots,{v}_m)$ 
over $\mathbb C$ with the fundamental commutation relation 
$[{u}_i,{v}_j]=-i\h\delta_{ij}$, where $\h$ is a positive constant.
The Heisenberg algebra $({\Cal H}_{2m}[\nu];*)$ is the algebra given 
by regarding the scalar parameter $\h$ in the Weyl algebra
${W}_{2m}[\h]$ to be a generator $\nu$ 
which commutes with all others. The difference of those two algebras
is seen in the next  
\begin{prop}\label{Heisenideal}
There is no nontrivial two-sided ideal of the Weyl algebra 
$(W_{2m}[\h];*)$. On the other hand, 
the Heisenberg algebra $({\mathcal H}_{2m}[\nu];*)$ has 
two-sided ideals corresponding to points of 
${\mathbb C}^{2m}$. 
\end{prop}

\noindent
{\bf Proof}\,\,is easy by observing the following: 
Suppose $\psi$ is a homomorphism of an algebra into 
${\mathbb C}$, and suppose $[x,y]_*=z$, then $\psi(z)=0$. 
It follows that there is no nontrivial two-sided ideal 
of the Weyl algebra $W_{2m}[\h]$. 

On the other hand $\nu{*}{\mathcal H}_{2m}[\nu]$ is a two-sided ideal of 
${\mathcal H}_{2m}[\nu]$ such that quotient algebra 
is the usual commutative polynomial ring ${\mathbb C}[\pmb u]$. 
It is easy to see that for every 
${\pmb a}\in {\mathbb C}^{2m}$, 
the two-sided ideal of ${\mathbb C}[\pmb u]$ generated by ${\pmb u}{-}{\pmb a}$ is 
pull back to give an nontrivial two-sided ideal of ${\mathcal H}_{2m}[\nu]$. 
\hfill $\Box$

Thus, the Heisenberg algebra may be treated in 
$(H\!ol({\mathbb C}^{2m+1}),*)$. These are seen in 
\cite{om3}, pp195-200, pp300-305.

\bigskip
However $\nu$ in the Heisenberg algebra is often regarded as 
an invertible element. In precisely, this means 
one may join the inverse $\nu^{-1}$ to ${\Cal H}_{2m}[\nu]$. 
In this case, the extended algebra ${\Cal H}_{2m}[\nu, \nu^{-1}]$
turns out to be isomorphic to ${W}_{2m}[\h]$.   
In spite of this, there is big difference in the notion of automorphisms.  

\bigskip
Consider a one parameter transformation 
$E_t:{\Cal H}_{2m}[\nu]\to {\Cal H}_{2m}[\nu]$ by 
setting on generators as  
$$
E_t(\nu)=e^{2t}\nu,\quad E_t(u_i)=e^{t}u_i,\quad E_t(v_i)=e^{t}v_i, \quad i=1,\dots,m, \quad
t\in{\mathbb R}.
$$
$E_t$ extends naturally to an automorphism of the Heisenberg algebra 
\begin{equation}\label{Et}
E_t: {\Cal H}_{2m}[\nu]\to {\Cal H}_{2m}[\nu],
\end{equation} 
 but this is  not an automorphism of the Weyl algebra $W_{2m}[\h]$, as
 $\h$ is not fixed. We refer $E_t$ to an {\bf expansive
   automorphism}.

\medskip
Consider the infinitesimal generator 
$D=\frac{d}{dt}E_t\Big|_{t=0}$ of \eqref{Et}. Obviously, 
$$
D(\nu)=2\nu, \quad D(u_i)=u_i,\quad D(v_i)=v_i,\quad i=1,\dots,m.
$$
Suppose now that $D$ is given by 
$\frac{1}{\nu}{\rm{ad}}(\frac{1}{i}\tau)$ in 
the ordinary shape of Hamiltonian mechanics   
using a certain (virtual) Hamiltonian $\tau$. 

We join the virtual element $\tau$ to the Heisenberg 
algebra ${\Cal H}_{2m}[\nu]$ by setting  
\begin{equation}\label{contact}
[i\tau,\nu]={-}2\nu^2,\quad [i\tau, u_i]={-}\nu{*}u_i,\quad i=1,\dots,2m.
\end{equation}
Though $\nu^{-1}$ is not an element of 
${\Cal H}_{2m}[\nu]$, we note $[\nu^{-1},\tau]{=}2i$.

In Heisenberg algebra, $\nu^{-1}$ is often viewed as the 
(independent universal) ``energy variable''. Then its canonical 
conjugate $\tau$ may be viewed as the ``universal time'' which might 
be fixed by the negotiation among the allover universe just like the
Greenwich mean time. 
 
However, the existence of such a canonical conjugate element is very
controversially relating to the {\bf time-energy uncertainty}, 
as one may see in \cite{Bu}.   
From a view point of differential geometry, the system \eqref{contact} is obtained as the 
deformation quantization of a local normal form of a contact structure. 
(See Proposition\,\ref{Localcood}.)

\medskip
In this note we are interested in the algebra consisting of 
$E_t$-invariant elements. Note that in quantum mechanics there is no
strict notion of restricting variables, while it is trivial in differential
geometry, where everything is considered on a patchwork of local
coordinate system. Thus, we regard this as the algebra of (non-commutative) unit sphere 
$S^{2m-1}{=}\{\sum_{k=1}^m(u_k^2{+}v_k^2){=}1\}$.   
This algebra is generated by 
\begin{equation}\label{nicecont}
\rho^{-2}{*}\nu, \,\,\rho^{-1}{*}{u}_1,\cdots,\rho^{-1}{*}{u}_m,\,\, 
\rho^{-1}{*}{v}_1,\cdots,\rho^{-1}{*}{v}_m
\end{equation}
where $\rho$ is an element satisfying $E_t\rho{=}e^t\rho$. 
Obviously, $\rho^{-1}$ is {\it not} an element of 
${\Cal H}_{2m}[\nu]$. Hence we have to join such an element to
the algebra ${\Cal H}_{2m}[\nu]$ to regulate the system, but the generated algebra depends   
how the commutation relations $[\rho^{-2}, \nu]$, 
$[\rho^{-1}, u_i]$, $[\rho^{-1}, v_j]$ are defined. If these are $0$, then 
the generated algebra is isomorphic to the Heisenberg algebra 
${\Cal H}_{2m}[\rho^{-2}{*}\nu]$, hence this procedure does not 
make the restriction of variables. 

\medskip 
The case that $\rho_*^2{=}\sum_{i=1}^{m}(u_i^2{+}v_i^2)$ may be the 
most familiar case. In physics, $i\nu$ and the transcendental element 
$\sqrt[*]{\sum_{i=1}^{m}(u_i^2{+}v_i^2)}$ are often treated as 
a selfadjoint semi-bounded operators.  
This implies that $\tau$ cannot be a self-ajoint operator. 
This is because as follows:

\medskip 
If it is possible, there must be a one parameter group of unitary operators 
$e_*^{it\tau}$, $t\in{\mathbb R}$. 
Suppose $e_*^{it\tau}$, $t\in{\mathbb R}$, exists by {\it itself} 
and $A_t{=}e_*^{it\tau}{*}\nu^{-1}{*}e_*^{-it\tau}$ is a welldefined 
real analytic group action in  $t$. 
We see that $A_t$ must
satisfy the equation 
$$
\frac{d}{dt}A_t{=}{\rm{ad}}(i\tau)A_t, \quad A_0{=}\nu^{-1}.
$$
Hence the uniqueness of the real analytic solution and 
the identity $[i \tau,\nu^{-1}]{=}2$ give 
$$
e_*^{it\tau}{*}\nu^{-1}{*}e_*^{-it\tau}{=}e^{t{\rm{ad}}(i\tau)}\nu^{-1}
{=}\nu^{-1}{+}2t{=}\nu^{-1}(1{+}2\nu t).
$$
It follows $\nu^{-1}{+}2t$ is invertible for every $t{\in}{\mathbb R}$.
This against the assumption $\nu$ is self-adjoint.

\medskip
On the other hand, as $\rho_*^{-2}{*}\nu$ is $E_t$-invariant, 
we have the identity 
$[\tau,\rho_*^{-2}{*}\nu]{=}0$, 
which easily yields 
$[\tau,\rho_*^{-2}]{*}\nu{-}\rho_*^{-2}2i{=}0$, 
hence  
$[i\tau,\rho_*^{-2}]{=}2\nu{*}\rho_*^{-2}$, 
$[i\tau,[i\tau,\rho_*^{-2}]]{=}0$, $\cdots$. 
Thus, the real analytic solution of the equation
$$
\frac{d}{ds}e_*^{is\tau}{*}\rho_*^{-2}{*}e_*^{-is\tau}{=}
{\rm{ad}}(i\tau)e_*^{is\tau}{*}\rho_*^{-2}{*}e_*^{-is\tau}
$$
is $e_*^{is\tau}{*}\rho_*^{-2}{*}e_*^{-is\tau}
{=}(1{+}2\nu s){*}\rho_*^{-2}$. 
It follows ${\rho}_*^{-2}{*}e_*^{-is\tau}{*}\rho_*^{2}
{=}(1{+}{2\nu s})e_*^{-is\tau}$. 
That is, $e_*^{-is\tau}$ is an eigenvector of  
${\rm{Ad}}(\rho^{-2})$.  
Hence in the situation that 
${\rm{Ad}}(\rho_*^{-2})$ has discrete eigenvectors,
we have to think $e_*^{ -is\tau}$ forms a discrete set. 
This may not be realistic.

\medskip
These observation suggest that $e_*^{it\tau}$ cannot
exist by {\it itself}. That is, the equation 
$$
\frac{d}{dt}F_t{=}i\tau{*}F_t, \quad F_0{=}f 
$$
may not be solved for the initial condition $1$ or even if it exists
solutions may not be unique.

%

\medskip
On the other hand in the theory of deformation 
quantization (cf.\cite{BF}), operator representations are  
required only in the final stage. In the beginning, we have only to 
extend ${W}_{2m}[\h]$ to a topological two-sided  
${W}_{2m}[\h]$-module. In an extended system, one may define 
$\sqrt[*]{\sum_{i=1}^{m}(u_i^2{+}v_i^2)}$ without using operator representations.
One has no need to care about positivity or self-adjointness, 
as there are no such notions. We are now free from the operator theory.
 
In this note, we propose an approach to the problem of the time-
energy uncertainty from a little more differential geometrical view
point together with the notion of abstract vacuums in the next section.  

\section{$\mu$-regulated algebra}\label{prelim2}
We begin with the notion which will be convenient to 
understand the motivation of the theory of deformation quantization (cf.\cite{BF}), and  
systems appeared by the restriction of the reduction 
procedures. Although it is not directly relevant to this note, 
$\mu$-regulated algebras are also very convenient to 
treat the calculus of pseudo-differential operators on manifolds (cf.\cite{om3}).
These will be discussed in forthcoming notes.

\bigskip
Let $\Cal A$ be a topological algebra with 1 over $\mathbb C$, which
may not be a complete topological space. 
We denote the product in $\Cal A$ by $a*b$.  
$\Cal A$ is called a $\mu$-{\it regulated algebra}, if there is 
an element $\mu\in \Cal A$, called a {\it regulator}, such that  
$(\Cal A,\mu)$ satisfies the following:  

\begin{description}
\item[(A.1)] $[\mu,\Cal A]\subset \mu\!*\!\Cal A \!*\!\mu$.
\item[(A.2)] $[\Cal A ,\Cal A]\subset\mu\!*\!\Cal A$.\quad($\Cal A$ is abelian modulo $\mu$.)
\item[(A.3)] There is a closed subspace $B$ such that  
          $\Cal A = B\oplus \mu\!*\!\Cal A$  $($topological direct sum$)$.
\item[(A.4)] Mappings 
       $\mu* : \Cal A \rightarrow \mu\!*\!\Cal A$,\,\,\,
               $*\mu : \Cal A\rightarrow \Cal A\!*\!\mu$
defined by $a\rightarrow \mu*a$, $a\rightarrow a*\mu$ respectively are 
linear isomorphisms.
\end{description}
\bigskip
 (A.1) shows 
$a{*}\mu{=}\mu{*}a{+}\mu{*}^{\exists}b{*}\mu{=}\mu{*}(a{+}b{*}\mu)$, hence
$\mu{*}{\mathcal A}\subset{\mathcal A}{*}\mu\subset\mu{*}{\mathcal A}$.
Thus $\mu{*}\Cal A{=}\Cal A{*}\mu$. It is a closed subspace, 
and hence $\mu{*}\Cal A$ is a closed two-sided ideal
of $\Cal A$. (A.2) shows the factor space $B=\Cal A/\mu{*}\Cal A$ is a
topological commutative algebra.  
Note that complementary space $B$ is not unique in general, but the quotient  
algebras are mutually isomorphic commutative algebras. 

\medskip
By the properties (A.3), (A.4), $\Cal A$ is decomposed for every $N$
into 
\begin{equation}\label{tag 0.4}
\Cal A = B\oplus\mu\!*\!B\oplus\cdots
    \oplus\mu^{N-1}\!*\!B\oplus \mu^N\!*\!\Cal A.
\end{equation}
We set $\Cal A^{-\infty}{=}\underset{k}{\bigcap}\,\,\mu^k\!*\!\Cal
A$. Then, $\mu\not\in \Cal A^{-\infty}$.
If $\Cal A^{-\infty}{=}\{0\}$, $\Cal A$ is said to be {\it analytic}
or {\it formal}. 

\medskip
For every $a, b\in B$, the decomposition \eqref{tag 0.4} gives    
\begin{equation}\label{tag 0.5}
a*b \sim
{\sum}_{k\geq 0}\mu^k *\pi_k(a,b), \quad \pi_k(a,b)\in B.
\end{equation}
We see $\pi_0(a,b)=a{\cdot}b$ is an associative commutative product, and  
the skew part $\pi^-_1$ of $\pi_1$ gives a biderivation of 
$B\times B$ into $B$. $\pi^-_1(a,b)$ is denoted by $\{a,b\}$ and
often called as Poisson bracket product.

\medskip
The property (A.4) permits us to join the inverse $\mu^{-1}$ 
to the algebra $\Cal A$. By setting 
$\mu{*}\mu^{-1}{=}\mu^{-1}{*}\mu{=}1$ and
$[\mu^{-1},a]{=}-\mu^{-1}{*}[\mu,a]{*}\mu^{-1}$, 
ad$(i\mu^{-1})$ gives a derivation of $\Cal A$, and it is decomposed as 
\begin{equation}\label{tag{3.6}}
{\rm{ad}}(i\mu^{-1})(a) = \xi_0(a)+\mu{*}\xi_{1}(a){+}\cdots
{+}\mu^k{*}\xi_k(a){+}\cdots,\quad a\in B
\end{equation}
where $\xi_0$ is a derivation of $(B,\cdot)$. Hence $\xi_0$ is
viewed as a vector field, which is called the {\it characteristic
  vector field}. The parameter of its integral curves is often viewed
as the ``time'' in differential geometry.  
Similarly, $(A.2)$ shows that 
ad$(a{*}\mu^{-1})$ is an outer derivation of $\Cal A$ for every
$a{\in}\Cal A$, and $(\Cal A{*}\mu^{-1}, [\,\,,\,\,])$ forms a Lie
algebra over $\mathbb C$ containing $(\Cal A, [\,\,,\,\,])$ as a Lie ideal. The 
quotient Lie algebra $\Cal A{*}\mu^{-1}/{\Cal A}$ is called the 
Jacobi algebra whose bracket product is  given by 
$$
\{f,g\}_{c}{=}f\xi_0(g){-}g\xi_0(f){+}\{f,g\}, \quad f, g\in B.
$$

\medskip
The enveloping associative algebra  of 
$(\Cal A{*}\mu^{-1}, [\,\,,\,\,])$ will be denoted by $\Cal A[\mu^{-1}]$.
This is often more important than $(\Cal A,*)$ itself or the Lie
algebra $(\Cal A{*}\mu^{-1}, [\,\,,\,\,])$. As elements of 
$\Cal A{*}\mu^{-1}$ are represented by unbounded operators in general, 
it seems that the complex Lie algebra $(\Cal A{*}\mu^{-1}, [\,\,,\,\,])$
cannot be a Lie algebra of a certain ``Lie group'' (cf.\cite{om3} pp.167-169). 

\medskip
Let $B_{0}{=}\{f{\in}B; \xi_0(f){=}0\}$. $B_{0}$ is a closed subalgebra of $(B,\cdot)$. 
There may be the case $B_{0}{=}B$, i.e. $\xi_0{=}0$. If this is the case, then 
$[\mu^{-1},B]\subset \mu{*}{\Cal A}$ and $\pi_1^-$ satisfies the Jacobi
identity, that is, $(B, \pi_{1}^{-})$ is a Lie algebra.
If $\xi_{0}{=}0$ and 
$\pi_1^-$ is nondegenerate on the space $B$, then
$(B,\cdot,\pi_1^-,\mu)$ is called an {\bf abstract symplectic  structure}.        
If $\xi_0{\not=}0$ and  $\pi_1^-$ is nondegenerate on the
space $B_{0}$, 
then $(B,\cdot,\xi_0, \pi_1^-)$ is called 
{\bf an abstract contact structure}.

In this context, Heisenberg
algebra is often regarded as a subalgebra of an abstract
contact algebra or a quantized contact algebra. 
In this case,  Heisenberg algebra appears as 
${\Cal A}_0{=}\{a{\in}{\Cal A};[\mu^{-1},a]=0\}$.

\bigskip
Although the choice of the complementary space $B$ is not unique,
but it depends on the  expression parameters
mentioned below, it is natural to think that all differential
geometrical structures defined on $(B,\cdot)$ should be expressed 
by notions involved in the first few terms of the $\mu$-series. 
This gives indeed the motivation of ``deformation quantization''.  
If it is possible, then such a geometrical
structure is called to be {\it deformation quantizable}. 

If this is the sole question, one has only to make a $\mu$-regulated
algebra where $\mu$-series are formal power series, and then the 
quantization problems would have almost been settled. Namely 
{\it every Poisson algebra is deformation quantizable} (cf. \cite{Ko}). 

However, it is natural to think that the theory via formal $\mu$ is 
not a true physics but only a probe toward the non-formal theory. 

\bigskip
In differential geometry, $(B,\cdot)$ is often assumed to 
be a commutative function algebra on a {\it phase}  
manifold $M$, and $\mu^{-1}$ is a real valued function on $M$, which
may be called {\it hamiltonian}.
  
As $\xi_0(\mu)=0$, the equality $\mu^{-1}{=}s$ may be regarded as the {\it energy surface} of 
level $s$, and $\mu^{-1}$ may be regarded as a local  
coordinate function on a neighborhood of the surface.  
A suitable local assumption permits us to consider a local coordinate function $\tau$
such that $\xi_0=\partial_{\tau}$, hence $\partial_{\tau}B_0{=}0$ and $\xi_0(\tau)=1$. 
Hence in differential geometry, a contact structure appears on the
energy surface and $\tau$ is involved as a coordinate function of the
configuration space (space-time). 
  
\medskip
In functional analysis, the eigenspace decomposition
$\sum_{s{\in}I}\oplus{\mathcal A}_{s}$ of 
$\rm{ad}(\mu^{-1}): {\mathcal A}\to{\mathcal A}$
is treated under suitable assumptions such as semi-boundedness or positivity. 
Let $\{a_{s}; s{\in}I\}$ be a family of eigenvectors, i.e. 
$[\mu^{-1},a_s]=sa_{s}$. Then, roughly speaking, we
have 
$$
[\mu^{-1},\log a_{s}]=s a_{s}/a_{s}=s.
$$
Hence, if $\tau_{s}{=}\log a_{s}$ exists, then the family   
$\{\tau_{s}; s{\in}I\}$ may be viewed as the canonical conjugate variable of 
$\mu^{-1}{=}\{\mu^{-1}; s{\in}I\}$ by regarding every
$f{\in}{\Cal A}$ as a constant family $\{f; s{\in}I\}$.

A $\mu$-regulated algebra with an abstract contact structure is 
called a {\bf contact Weyl algebra}. Such an algebra often appears 
in the shape ${\Cal A}_0[\tau]$.
In general, there is a serious unrepairable gap 
between these two view points. Main reason of this gap is 
that local structures are leading target in differential geometry, but 
the global structures are concerned in analysis. One of the reason is
that the method of obtaining {\it normal form} is given only in the
formal level, as it will be mentioned below.  

\bigskip
\noindent
{\bf Localization theorem}.\,\,If a $\mu$-regulated algebra   
$({\mathcal A},\mu)$ is defined by setting a $*$-product on a function space of a 
manifold $M$, then by choosing a 
suitable complementary subspace $B$, all $\xi_k$ and $\pi_k$ become 
respectively differential operators and bidifferential operators.   
This is called the localization theorem proved in \cite{om3}. 
Note that ${\Cal A}/{\Cal A}^{-\infty}$ is a (formal) $\mu$-regulated 
algebra. The localization theorem shows that ${\Cal A}/{\Cal A}^{-\infty}$ 
can be restricted on every open subset $U$ of $M$ by using the restrictions
of $\pi_k$ and $\xi_k$ on $U$. 

\bigskip
\noindent
{\bf Quantum Darboux theorem} (cf.\cite{om3}).\, If the abstract symplectic
structure is involved further, then the localization theorem  
and stepwise repairing of coordinate functions 
show that there is an open neighborhood $U_p$ of $p{\in}M$ on which 
${\Cal A}/{\Cal A}^{-\infty}$ is expressed as a Weyl algebra. 
(See the next section for the definition of the Weyl algebra.) 

\bigskip 
If such a localization is obtained by joining ``divisors'', then 
one may not loose the information about ${\Cal A}^{-\infty}$.
Note also that the $\mu$-regulated algebra $\widetilde{\Cal A}$ reconstructed 
by localized generator system must have $\widetilde{\Cal A}^{-\infty}$ part. 
That is, localized system might not be formal. Now it arises a big, basic question
how $\widetilde{\Cal A}$ relates to the original ${\Cal A}$. A system
in a laboratory must be a certain localization of the system of the whole universe.

\medskip
The next Proposition shows that \eqref{contact} is
a standard local coordinate system.
\begin{prop}\label{Localcood}
Suppose $(B, \cdot)$ is the space of $C^{\infty}$ functions on a manifold $M$
and $\mu$ is given as a smooth function on $M$.
Then, there is a local coordinate system with the property
\eqref{contact} on a tubular neighborhood of a compact 
fragment of an orbit of $\xi_0$.
\end{prop}

\noindent
{\bf Proof}\,(Sketch)\,\,  As $[\mu^{-1},\mu]{=}0$ gives $\xi_0(\mu)=0$, every orbit of 
$\xi_0$ is sitting on a subspace $\mu{=}s$(constant).  One may assume
$\mu{=}s$ defines a manifold $M_{s}$ of codimension one.  
Consider a compact fragment $c(\tau)$ of an orbit of $\xi_0$. We assume
$\xi_0(\tau)=1$, and we take a system of local
coordinate $s, u_1,\cdots,u_{m-1},$ $v_1,\cdots,v_{m-1}$ 
on a disk transversal to $c(*)$. It is clear that $\xi_0(s)=0$ and
one may assume  
$\xi_0(u_i)=0$ and $\xi_0(v_i)=0$ for $i=1\sim m-1$. 
As $\pi_1^-$ is nondegenerate, classical Darboux
theorem shows that one may assume $\pi_1^{-}(u_i,v_j){=}\delta_{ij}$.

By the localization theorem, the original $*$-product can be 
restricted to the functions on this open
neighborhood. Hence, we have $\mu{=}s$ and
$$
[\mu^{-1}, u_k]{=}0,\quad mod(\mu),\quad
[\mu^{-1},v_k]{=}0,\quad mod(\mu), \quad [u_i,v_j]=\mu i\delta_{ij}
\quad mod(\mu^2)
$$
On the other hand, we denote 
$$
[\begin{bmatrix}
 {\pmb u}\\{\pmb v}
 \end{bmatrix},
({}^t\!{\pmb u},{}^t\!{\pmb v})]=
2i\mu^2\begin{bmatrix}
0&-I\\
I&0
\end{bmatrix},\quad 
{\rm{ad}}(\tau)
\begin{bmatrix}
 {\pmb u}\\{\pmb v}
 \end{bmatrix}{=}
A\begin{bmatrix}
 {\pmb u}\\{\pmb v}
 \end{bmatrix},
\quad 
{\rm{ad}}(\tau)({}^t\!{\pmb u},{}^t\!{\pmb v})=({}^t\!{\pmb u},{}^t\!{\pmb v}){}^t\!A.
$$
The identity 
$$
2i\mu^2\delta_{ij}{=}[\tau,\mu\delta_{ij}]=[\tau, [u_i,v_j]]=[[\tau,u_i],v_j]{+}[u_i,[\tau,v_j]]
$$
together with $[\tau, [u_i,u_j]]=[\tau, [v_i,v_j]]=0$ gives 
$2i\mu^2J=AJ{+}J{}^t\!A$. It follows 
$$
(A{-}i\mu)J{+}J{}^t(\!A{-}i\mu)=0,\quad i.e.\,\,\, (A{-}i\mu)J
\,\,\,\text{is a symmetric matrix}.
$$
By a suitable linear change of coordinate one may set 
that 
$$
[\tau,u_k]{=}a_k\mu{*}u_k{+}\cdots,\quad 
[\tau,v_k]{=}b_k\mu{*}v_k{+}\cdots,
$$
and one may assume $a_k=b_k=i$ after a further tuning.
Suitable change of coordinate by a standard linearization procedure 
gives that $[\tau,u_i]{=}i\mu{*}u_i$, $[\tau,v_i]{=}i\mu{*}v_i$. 
Embed this system into a symplectic manifold, and  
the quantum Darboux theorem gives the desired result.
   ${}$ \hfill $\Box$

\subsection{Abstract vacuums}\label{sec:absvac}
In physics, the vacuum may be defined as the 
 lowest eigenstate of energy, but in our situation 
${\rm{ad}}(\mu^{-1})$ may not be semi-bounded. 
Here, we propose a little different notion of vacuums. We call  
an idempotent element $\varpi{\in}{\Cal A}$ 
an {\bf abstract vacuum} if 
\begin{equation}\label{absvacuum}
\varpi{*}\varpi{=}\varpi, \quad 
\varpi{*}{\Cal A}{*}\varpi{=}{\mathbb C}{*}\varpi.
\end{equation}
Such an element is not contained in the Weyl algebra, but there are
a lot of such elements in a transcendentally extended Weyl algebra. 
If $\varpi$ is an abstract vacuum and $\varpi{*}g{*}\varpi{=}\lambda_g\varpi$,
$\lambda_g\in{\mathbb C}{\setminus}\{0\}$, then  
$\lambda_g^{-1}g{*}\varpi$ is an abstract vacuum. 

\bigskip

It is remarkable that an abstract vacuum exists only in ${\Cal  A}^{-\infty}$, 
i.e  ${\Cal  A}$ non-formal . 
This is proved as follows: 
If $\varpi{*}\mu{*}\varpi{=}c\varpi$, $c\not=0$, then 
$\frac{1}{c}\mu{*}\varpi{\in}\mu{*}{\Cal A}$ is an abstract vacuum. As
this is idempotent, we see $\frac{1}{c}\mu{*}\varpi{\in}{\Cal A}^{-\infty}$   
and hence $\varpi{\in}{\Cal A}^{-\infty}$. If $c{=}0$, then $(A.1)$
gives $0{=}\varpi{*}\mu{*}\varpi{=}\mu{*}\varpi{+}\mu{*}a{*}\mu{*}\varpi$. Hence,
$\mu{*}\varpi{\in}\mu^2{*}{\Cal A}$, and then 
$\varpi{\in}\mu{*}{\Cal A}$. 
It follows $\varpi{\in}{\Cal A}^{-\infty}$, 
and ${\Cal A}{*}\varpi{*}{\Cal A}\subset {\Cal A}^{-\infty}$.

\medskip
If ${\Cal A}$ is generated by $1, x_1,\cdots, x_n$, and $x_i{*}\varpi{=}0$ 
or  $\varpi{*}x_i{=}0$ holds for every $i$, then $\varpi$ is an 
abstract vacuum. We call such an abstract vacuum a 
{\bf standard abstract vacuum}. (Note that there is a non-standard
abstract vacuum, called a pseudo-vacuum in \cite{ommy6}.)

Sometimes, it is natural to assume 
\begin{equation}\label{assume1}
\mu{*}\varpi{=}c\varpi,\quad c>0.
\end{equation}
If the eigenspace decomposition
$\sum_{s{\in}I}\oplus{\mathcal A}_{s}$ of 
$\rm{ad}(\mu^{-1})$ is given, then 
setting ${\Cal L}_s{=}{\Cal A}_s/L\cap{\Cal A}_s$, we have 
$$
\mu^{-1}{*}\phi_s{*}\varpi=(s{+}c^{-1})\phi_s{*}\varpi, 
\quad \mu{*}\phi_s{*}\varpi=\frac{c}{cs{+}1}\phi_s{*}\varpi, \quad 
\phi_s\in{\Cal L}_s.  
$$
As $\mu^{-1}$ is viewed as the energy, the positivity $s{+}c^{-1}{>}0$ may
be required, but there is no effective relation between $\varpi$ and 
$\phi_s$. 

\begin{center}
\fbox{
\parbox{.8\linewidth}
{The observation above suggests that the next target of the theory of
deformation quantization is to construct $\mu$-regulated algebras with 
abstract vacuums.}}
\end{center}

\medskip
As $1{=}\varpi{+}(1{-}\varpi)$ and ${\Cal A}{=}{\Cal A}{*}\varpi\oplus{\Cal A}{*}(1{-}\varpi)$,  
we see 
$$
\Cal A{=}{\mathbb C}\varpi\oplus
    (1{-}\varpi){*}{\Cal A}{*}\varpi\oplus
     \varpi{*}{\Cal A}{*}(1{-}\varpi)\oplus (1{-}\varpi){*}{\Cal A}(1{-}\varpi).  
$$

In what follows we assume there exist closed linear subspaces  
${\Cal L}\subset(1{-}\varpi){*}{\Cal A}$ and ${\Cal R}\subset{\Cal
  A}{*}(1{-}\varpi)$ such that  
${\Cal L}\cap{\Cal A}^{-\infty}{=}\{0\}$,  
${\Cal R}\cap{\Cal A}^{-\infty}{=}\{0\}$  and 
$$
(1{-}\varpi){*}{\Cal L}{*}\varpi{=}(1{-}\varpi){*}{\Cal A}{*}\varpi,
\quad  
\varpi{*}{\Cal R}{*}(1{-}\varpi){=}\varpi{*}{\Cal A}{*}(1{-}\varpi)
$$
and the projections 
$$
{\pi}: {\Cal L}{=}(1{-}\varpi){*}{\Cal L}\to (1{-}\varpi){*}{\Cal L}{*}\varpi,\quad 
{\pi}: {\Cal R}{=}{\Cal R}{*}(1{-}\varpi)\to \varpi{*}{\Cal R}{*}(1{-}\varpi)
$$
are linear isomorphic. We set 
$$
\tilde{\Cal L}{=}{\mathbb C}{*}1\oplus(1{-}\varpi){*}{\Cal L},\quad 
\tilde{\Cal R}{=}{\mathbb C}{*}1\oplus{\Cal R}{*}(1{-}\varpi).
$$
Note that the projections 
$\pi:\tilde{\Cal L}\to \tilde{\Cal L}{*}\varpi$,\,\,
$\pi:\tilde{\Cal R}\to \varpi{*}\tilde{\Cal R}$ are linear
isomorphisms.  
As every $a{\in}{\Cal A}$ is written by 
$$
a{=}\varpi{*}a{*}\varpi{+}(1{-}\varpi){*}a{*}\varpi{+}
    \varpi{*}a{*}(1{-}\varpi){+}(1{-}\varpi){*}a{*}(1{-}\varpi), \quad 
\varpi{*}a{*}\varpi{=}\delta_a\varpi{\in}{\mathbb C}{\varpi},
$$ 
we see that $\delta_a\varpi{+}(1{-}\varpi){*}a{*}\varpi\in\tilde{\Cal L}{*}\varpi$, 
and $\delta_a{+}(1{-}\varpi){*}a\in \tilde{\Cal L}$.
\begin{prop}\label{identify}
Let $L=\{f\in{\Cal A}; f{*}\varpi{=}0\}$. Then 
$L=\varpi{*}{\Cal A}{*}(1{-}\varpi){\oplus} 
(1{-}\varpi){*}{\Cal A}(1{-}\varpi)$ 
and 
${\Cal A}/L$ is linearly isomorphic to $\tilde{\Cal L}$ and
$\tilde{\Cal L}{*}\varpi$. An element $a$ of ${\Cal L}$ is characterized
by $a{=}(1{-}\varpi){*}a$. 

Similarly,  
let $R=\{f\in{\Cal A}; {\varpi}{*}f{=}0\}$. Then 
$R=(1{-}\varpi){*}{\Cal A}{\oplus}(1{-}\varpi){*}{\Cal A}(1{-}\varpi)$ 
and 
$R{\setminus}{\Cal A}$ is linearly isomorphic to $\tilde{\Cal R}$ and
$\varpi{*}\tilde{\Cal L}$. 
An element $b$ of ${\Cal R}$ is characterized
by $b{=}b{*}(1{-}\varpi)$. 
\end{prop}
The regular representation, which is called often a 
{\bf vacuum representation}  
of ${\Cal A}$ is defined for $a{\in}{\Cal A}$ by  
$$
P(a): \tilde{\Cal L}\to\tilde{\Cal L},\quad P(a)(\phi){=}\pi^{-1}(a{*}\phi{*}\varpi).
$$
Since $\pi^{-1}(a{*}\varpi){*}\varpi{=}a{*}\varpi$, it is clear that 
$P(a)P(b){=}P(a{*}b)$. In particular,     
\begin{equation}\label{deltafunc}
P(\varpi)(\phi){=}\pi^{-1}(\varpi{*}\phi{*}\varpi){=}\delta_{\phi},
\quad \delta_{\phi}{\in}{\mathbb C},\quad \delta_{\varpi}{=}1
\end{equation}
is a projection operator of rank one.  
In particular, 
$\tilde{\Cal L}{*}\tilde{\Cal L}{*}\varpi{=}\tilde{\Cal L}{*}\varpi$, 
but $\tilde{\Cal L}{*}\tilde{\Cal L}{=}\tilde{\Cal L}$ does not hold
in general.

\medskip
On the other hand, there is a natural bilinear form  
$\langle\,\,,\,\,\rangle_{\varpi}: R{\setminus}{\Cal A}\times{\Cal A}/L\to {\mathbb C}$ 
over ${\mathbb C}$ defined by 
\begin{equation}\label{bilinear}
\varpi{*}\psi{*}\phi{*}\varpi{=}\langle\psi{*}\phi\rangle_{\varpi}\varpi.
\end{equation}
This is identified with the bilinear form  
\begin{equation}\label{bilinear}
\langle\,\,,\,\,\rangle_{\varpi}:
 \tilde{\Cal R}\times \tilde{\Cal L}\to  {\mathbb C}.
\end{equation}
In this note, we mainly consider the matrix representation given in the form
$\tilde{\Cal L}{*}\varpi{*}\tilde{\Cal R}$. Obviously, we have 
$\tilde{\Cal L}{*}\varpi{*}\tilde{\Cal R}\subset{\Cal A}^{-\infty}$
and  
$$
(\tilde{\Cal L}{*}\varpi{*}\tilde{\Cal R}){*}
(\tilde{\Cal L}{*}\varpi{*}\tilde{\Cal R}){=}
 \tilde{\Cal L}{*}\varpi{*}\tilde{\Cal R}.
$$

In typical examples, the bilinear form \eqref{bilinear}
is non-degenerate, and there are linear basis 
$$
\tilde{\Cal L}: e_0,  e_1, e_2, \cdots, e_k,\cdots,\quad 
\tilde{\Cal R}: f_0,  f_1, f_2, \cdots, f_k, \cdots,\quad e_0{=}1{=}f_0, 
$$
such that $\langle e_i,f_j\rangle_{\varpi}{=}\delta_{ij}$. In this
case, we see 
$e_k{*}\varpi{*}f_{\ell}$ is the $(k,\ell)$ matrix element and 
$\varpi{=}e_0{*}\varpi{*}f_{0}$. 
They are elements of ${\Cal  A}^{-\infty}$, but 
the representation space is spanned by 
$\{e_k{*}\varpi; k=0,1,2,\cdots\}$ together with 
certain topologies.

\section{Extended Weyl algebra} 

Elements such as abstract vacuums are transcendental elements. 
For the systematic treatment for such elements, we have to
begin with the transcendental extension of Weyl algebra. 

Let $\pmb u{=}(u_1,\cdots,u_m,u_{m+1},\cdots,u_{2m})$. 
For an arbitrary fixed $2m{\times}2m$-complex symmetric matrix 
$K{\in}{\mathcal S}_{\mathbb C}(2m)$, we set $\Lambda{=}K{+}J$ where
$J$ 
is the standard skew-symmetric matrix  
$J{=}\tiny{
\begin{bmatrix}
0&-I_m\\
I_m&0
\end{bmatrix}}$. 
We define a product ${*}_{_{K}}$ on the space
of polynomials  ${\mathbb C}[\pmb u]$ by the formula 
\begin{equation}
 \label{eq:KK}
 f*_{_{K}}g=fe^{\frac{i\h}{2}
(\sum\overleftarrow{\partial_{u_i}}
{\Lambda}{}^{ij}\overrightarrow{\partial_{u_j}})}g
=\sum_{k}\frac{(i\h)^k}{k!2^k}
{\Lambda}^{i_1j_1}\!{\cdots}{\Lambda}^{i_kj_k}
\partial_{u_{i_1}}\!{\cdots}\partial_{u_{i_k}}f\,\,
\partial_{u_{j_1}}\!{\cdots}\partial_{u_{j_k}}g, \quad 
\end{equation}
where $i\h\in {\mathbb C}{\setminus}\{0\}$ is a complex parameter. When $K{=}0$, the product formula
\eqref{eq:KK} is called the {\bf Moyal product formula}.    
(In  \cite{OMMY7}, \cite{OMMY3}, \cite{OMMY4},
\cite{OMMY5}, \cite{ommy6}, 
$*_{_K}$-product is first denoted by $*_{_{\Lambda}}$, or $*_{_{K{+}J}}$, and
then the notation $*_{_K}$ is used after fixing the skew part $J$.)

It is known and not hard to prove that 
$({\mathbb C}[\pmb u],*_{_{K}})$ are mutually 
isomorphic associative algebras for every $K$. 
The Weyl algebra $(W_{2m}[\h]; *)$ is the name of 
the isomorphism class. 

In calculations in the algebra where we need only 
the commutation relations,  we have no need to use the expression
by $K$. Thus, it is natural to think that the algebraic essence must be free from
their expressions, which may be called {\it the independence of
  ordering principle} (IOP in short). But this is a dangerous trip. 
In highly transcendental elements, the expression parameter $K$ plays 
an essential role (cf.\S,\,\ref{comments}). 
We have seen in \cite{OMMY3} that the expression parameter is 
essential even in the one variable commutative case.
 
\medskip
If $K$ is fixed, the $*_{_K}$-product formula gives 
a way of univalent expression for elements of $(W_{2m}[\h]; *)$. 
Every element $f_*{\in}(W_{2m}[\h]; *)$ 
is expressed in the form of ordinary polynomial, which we 
denote by ${:}f_*{:}_{_K}\in {\mathbb C}[\pmb u]$ and call this 
{\bf the $K$-ordered expression (or $K$-expression in brief) of} $f_*$.
In this context, the product formula \eqref{eq:KK} is often called a 
$K$-ordered expression. 
$K=(K^{ij}){=}(\langle u_i,Ku_j\rangle)$ is called an expression parameter.  
Via univalent expressions, one can consider topological completion of 
the algebra, and various transcendental elements.

\bigskip
The intertwiner between $K$-ordered expression and $K'$-ordered expression 
is explicitly given as follows:
\begin{prop}
\label{intwn}
For every $K, K'\in{\mathcal S}_{\mathbb C}(2m)$, the intertwiner
is defined by  
\begin{equation}
\label{intertwiner00}
I_{_K}^{^{K'}}(f)=
\exp\Big(\frac{i\h}{4}\sum_{i,j}(K^{'ij}{-}K^{ij})
\partial_{u_i}\partial_{u_j}\Big)f \,\,
(=I_{0}^{^{K'}}(I_{0}^{^{K}})^{-1}(f)), 
\end{equation}
which gives an isomorphism 
$I_{_K}^{^{K'}}:({\mathbb C}[{\pmb u}]; *_{_{K}})\rightarrow 
({\mathbb C}[{\pmb u}]; *_{_{K'}})$.
Namely, 
for any $f,g \in {\mathbb C}[{\pmb u}],$ we have  
\begin{equation}\label{intertwiner2}
I_{_K}^{^{K'}}(f*_{_K}g)=
I_{_K}^{^{K'}}(f)*_{_{K'}}I_{_K}^{^{K'}}(g).
\end{equation}
\end{prop}
Intertwiners do not change the algebraic structure $*$, 
but these change the expression of elements by the ordinary 
commutative structure.  

The next formula is trivial, but often very useful in the 
concrete calculation: (Cf.\S\,\ref{comment2}.) 
Suppose $K=K_1{+}K_2$. Then 
\begin{equation}\label{trivial}
I_{_{K_2}}^{^{K_1{+}K_2}}I_0^{^{K_2}}=
I_0^{^{K_1{+}K_2}}.
\end{equation}

\bigskip
\noindent
{\bf Example 1} \,\,
Let $H\!ol({\mathbb C}^{2m})$ (resp. $C^{\infty}({\mathbb R}^{2m})$) be the space of
all holomorphic (resp. $C^{\infty}$-) functions on ${\mathbb C}^{2m}$
(resp. ${\mathbb R}^{2m})$.  
By setting $\nu=\h$ in \eqref{eq:KK}, and by fixing the expression parameter $K$ in
${\mathcal S}_{\mathbb C}(2m)$, we see that the Heisenberg algebra
${\Cal H}_{2m}[\nu]$ is a $\nu$-regulated algebra. 

Let  
${:}{\Cal A}{:}_{_K}=H\!ol({\mathbb C}^{2m})[[\nu]]$,  
(resp, ${:}{\Cal A}{:}_{_K}=C^{\infty}({\mathbb R}^{2m})[[\nu]]$),  
 the space of all formal power series of $\nu$. 
Then, \eqref{eq:KK} makes $({:}{\Cal A}{:}_{_K}; *_{_K})$ an associative algebra.
As they are all mutually isomorphic by the intertwiners, the
isomorphism class will be denoted by  $({\Cal A}; *)$.  
This is a $\nu$-regulated algebra such that $[\nu, {\Cal A}]{=}\{0\}$
and 
$$
\Cal A{=}H\!ol({\mathbb C}^{2m}){\oplus}\nu{*}{\Cal A}, \quad
{resp.}\quad 
\Cal A{=}C^{\infty}({\mathbb R}^{2m}){\oplus}\nu{*}{\Cal A},
$$ 
involving abstract symplectic structure. The isomorphism class  $({\Cal A}; *)$
will be called the {\bf formally extended Weyl algebra}, or 
the {\bf formally extended Heisenberg algebra}.

Although the proof will be skipped, the next one is an important
remark:
\begin{prop} 
Changing expression parameter $K$ corresponds to the replacement of complementary subspace $B$. 
In fact, the latter is much wider than changing $K$.  
\end{prop}

\subsection{Extension of products}\label{Extprodpp}
For every positive real number $p$, we set 
\begin{equation}
\label{sysnorm1}
  {\mathcal E}_p({\Bbb C}^{2m})=
\{f \in H\!ol({\Bbb C}^{2m})\, ;\, 
 \|f\|_{p,s}=\sup\, |f|\, e^{-s|{\pmb u}|^p}<\infty,\,\,\forall s>0\}
\end{equation}
where ${\pmb u}=(u_1,\ldots,u_{2m})$ and $|{\pmb u}|=(\sum_i|u_i|^2)^{1/2}$. 
The family of seminorms 
$\{||\,\cdot\,||_{p,s}\}_{s>0}$ induces a topology 
on ${\mathcal E}_p({\Bbb C}^{2m})$ and 
$({\mathcal E}_p({\Bbb C}^{2m} ),\cdot)$ is 
an associative commutative Fr\'{e}chet algebra, 
where the dot $\cdot$ is 
the ordinary product for functions in 
${\mathcal E}_p({\Bbb C}^{2m})$. 
It is easily seen that for $0<p<p'$, there is a 
continuous embedding 
\begin{equation}
  {\mathcal E}_p({\Bbb C}^{2m} )\subset 
{\mathcal E}_{p'}({\Bbb C}^{2m} ) 
\end{equation}   
as commutative Fr\'{e}chet algebras (cf.\,\cite{GS}), and that 
${\mathcal E}_p({\Bbb C}^{2m})$ is $G\!L(n,{\mathbb C})$-invariant. 

We denote  
\begin{equation}
{\mathcal E}_{p+}({\Bbb C}^{2m})
=\bigcap_{p'>p}{\mathcal E}_{p'}({\Bbb C}^{2m}),\quad 
(\text{with the intersection topology})
\end{equation}
It is obvious that every polynomial is contained in 
${\mathcal E}_{p}({\Bbb C}^{2m})$, that is 
$p({\pmb u})\in {\mathcal E}_{0+}({\Bbb C}^{2m})$, 
 and $\Bbb C[{\pmb u}]$ is dense 
in ${\mathcal E}_{p}({\Bbb C}^{2m})$ for any $p>0$ 
in the Fr{\'e}chet topology defined by the family of 
seminorms $\{||\,\,||_{p,s}\}_{s>0}$. 

\bigskip

We easily see that 
$e^{\sum a_iu_i}\in{\mathcal E}_{1+}({\Bbb C}^{2m})$. Moreover, it is not 
difficult to show that an exponential 
function $e^{p({\pmb u})}$ of a polynomial $p({\pmb u})$ of degree $d$ is 
contained in ${\mathcal E}_{d+}({\Bbb C}^{2m})$, but not in 
${\mathcal E}_{d}({\Bbb C}^{2m})$.

\begin{thm}\label{main01} $(Cf.\text{\rm{\cite{OMMY1},\cite{o-el3}}})$
For $0<p\leq 2$, the product formula \eqref{eq:KK} 
extends to give the following{:}

\noindent
$(1)$ The space $({\mathcal E}_{p}({\Bbb C}^{2m}),*_{_{K}})$ 
       forms a complete noncommutative topological associative
       algebra.

\noindent
$(2)$ The intertwiner $I_{_K}^{^{K'}}$ extends to give an 
isomorphism of  $({\mathcal E}_{p}({\Bbb C}^{2m}),*_{_{K}})$ 
onto $({\mathcal E}_{p}({\Bbb C}^{2m}),*_{_{K'}})$.
\end{thm}
Hence we denote the isomorphism class by 
$({\mathcal E}_{p}({\Bbb C}^{2m}),*)$ ($0\leq p\leq 2$), and we call
this the {\it extended Weyl algebra}. 
This is naturally an $\h$-regulated algebra. An element $H_*$ of 
$({\mathcal E}_{p}({\Bbb C}^{2m}),*)$ is a family 
$\{H_{_K}; K\in {\mathcal S}_{\mathbb C}(2m)\}$ such that
$I_{_K}^{^{K'}}H_{_K}{=}H_{_{K'}}$, but it is convenient to denote 
$H_{_K}$ by ${:}H_*{:}_{_K}$. 

\medskip
Note that the exponential function of a quadratic form can not be  
treated by Theorem\,\ref{main01}. 

\medskip
In the case $p{>}2$, we see the next one:
\begin{thm}\label{main02} $(Cf.\text{\rm{\cite{OMMY1},\cite{o-el3}}})$ 
For $p{>}2$, find $p'$ such that $\frac{1}{p}+\frac{1}{p'}{\geq}1$,
then the product \eqref{eq:KK}extends to a continuous bilinear mapping  
\begin{equation}
{\mathcal E}_{p}({\Bbb C}^{2m})\times
{\mathcal E}_{p'}({\Bbb C}^{2m})\rightarrow 
{\mathcal E}_{p}({\Bbb C}^{2m}),\quad 
{\mathcal E}_{p'}({\Bbb C}^{2m})\times
{\mathcal E}_{p}({\Bbb C}^{2m})\rightarrow 
{\mathcal E}_{p}({\Bbb C}^{2m}).  
\end{equation}
If two of $f, g, h\in{\mathcal E}_{p}({\Bbb C}^{2m})$ $(p{>}2)$ are in  
${\mathcal E}_{p'}({\Bbb C}^{2m})$, then the associativity 
$(f{*}_{_{K}}g){*}_{_{K}}h=f{*}_{_{K}}(g{*}_{_{K}}h)$ 
holds. Namely, ${\mathcal E}_{p}({\Bbb C}^{2m})$ is a two-sided 
${\mathcal E}_{p'}({\Bbb C}^{2m})$-module.
\end{thm}

\subsubsection{Several remarks  on the notation ${:}A_*{:}_{_K}$}\label{comments} 
In previous notes
we had used notations such as ${:}A_*{:}_{_K}$, and it will be used
often in what follows. Here we have to explain what $A_*$ is. 
This is in fact the family $\{A_{_K}; K\in{\mathcal I}\}$ 
where $A_{_K}$ are elements computed
by $K$-expression which are mutually intertwined by $I_{_{K}}^{^{K'}}$ 
for every $K, K'\in {\mathcal I}$. ${\mathcal I}$ is a subset of 
${\mathcal S}_{\mathbb C}(2m)$. If ${\mathcal I}$ contains an open
dense subset of ${\mathcal S}_{\mathbb C}(2m)$, then we say that
$A_*$ is given by a generic ordered expression. 

Sometimes, elements such as $A_*^{-1}$ or $\sum_{k\in{\mathbb Z}}e_*^{kA}$, 
${\mathcal I}$ is not a dense subset. 
It contains only a nonempty open subset. Even in such a
case we use notations ${:}A_*{:}_{_K}$ by indicating the domain ${\mathcal I}$ 
for $A_*$. 
 
\medskip 
It is sometimes very convenient to compute under a fixed expression
parameter $K$. In particular, Weyl ordered ($K=0$) expression and 
the normal ordered 
($K=
\left[
\begin{smallmatrix}
0&1\\
1&0\\
\end{smallmatrix}
\right]
$) expression are very convenient to compute. However, such
expressions are not generic. Therefore, the results might not be
a generic result. 

However, the next Proposition shows that such a restricted expression 
can be a useful tool.
\begin{prop}\label{judge}
Suppose we have an identity 
${:}C_*{:}_{_{K_0}}{=}{:}A_*{:}_{_{K_0}}{*_{_{K_0}}}{:}B_*{:}_{_{K_0}}$ under
a fixed $K_0$, and suppose ${:}A_*{:}_{_{K_t}}$, ${:}B_*{:}_{_{K_t}}$
and their ${*_{_{K_t}}}$ product 
${:}A_*{:}_{_{K_t}}{*_{_{K_t}}}{:}B_*{:}_{_{K_t}}$
are defined on a path $K_t$, $t{\in}[0,1]$. 
Then, one may define ${:}C_*{:}_{_{K_1}}$ by  
${:}C_*{:}_{_{K_1}}{=}{:}A_*{:}_{_{K_1}}{*_{_{K_1}}}{:}B_*{:}_{_{K_1}}$. 
\end{prop}
In general, the r.h.s. depends on the path to $K_0$ to $K_1$. Indeed,
it causes the double valued nature of the $*$-exponential functions of 
quadratic forms. 
Thus, a transcendental element $A_*$ in a $*$-algebra is defined always with
expression parameters ${\Cal I}(A_*)$, and sometimes with the path
from the reference point (cf. the notion of synchronized path
selection in \cite{OMMY5}). 

If ${\Cal I}(A_*)$ contains
open dense subset, then it may not be explicitly indicated. 
However, ${\Cal I}(A_*)$ might be restricted in a very small region.
Furthermore, in \cite{OMMY4}, \cite{ommy6}, we saw that 
a certain family of elements behaves very different way under some 
special expression parameters.

\subsection{Star-exponential functions}\label{starexp00}

The $*$-exponential function
$e_*^{tH_*}$ for $H_*{\in}(\Cal A,{*})$ is defined as the collection 
${:}e_*^{tH_*}{:}_{_K}$ of its $K$-expressions. If $\h$ is a formal
parameter, then  ${:}e_*^{tH_*}{:}_{_K}$ is defined as a series 
$\sum\frac{t^n}{n!}{:}H_*^n{:}_{_K}$ formally rearranged as power
series of $\h$. 
In the case $\h$ is not formal, the $*$-exponential function
$e_*^{tH_*}$ is not defined by a power series. If $H_*$ is a
polynomial of degree 3, then the radius of convergence is $0$ in general (cf. \cite{OMMY3}).

Instead, these are considered as 
the solution of an evolution equation
$$
\frac{d}{dt}{:}f_t{:}_{_K}{=}{:}H_*{:}_{_K}{*_{_K}}{:}f_t{:}_{_K}, \quad {:}f_0{:}_{_K}{=}1. 
$$
This is a differential equation, if $H_*$ is a polynomial, but the
solution may not exist in general, but a real analytic solution is
unique. 
The solution (if exists uniquely) with initial data ${:}g_*{:}_{_K}$ will be denoted by 
${:}e_*^{tH_*}{*}g_*{:}_{_K}$. 

\medskip
Note however even though ${:}e_*^{tH_*}$ is not defined by itself, there is a
case that ${:}e_*^{tH_*}{*}g_*{:}_{_K}$ is welldefined for some
initial data. Moreover, even
though ${:}e_*^{tH_*}{*}g_*{:}_{_K}$ is not defined, there is a case 
where ${:}e_*^{tH_*}{*}g_*{*}e_*^{-tH_*}{:}_{_K}$ is welldefined as 
$e^{t{\rm{ad}}(H_*)}g_*$.

To simplify notations, we often denote 
$\langle{\pmb a}{\varGamma},{\pmb b}\rangle=\sum_{ij=1}^{2m}{\varGamma}^{ij}a_ib_j$, 
$\langle{\pmb a},{\pmb u}\rangle=\sum_{i=1}^{2m} a_iu^i$. These will
be denoted also by 
${\pmb a}{\varGamma}\,{}^t{\pmb b}$ and 
$\langle{\pmb a},{\pmb u}\rangle={\pmb a}\,{}^t\!{\pmb u}$. It is easy
to see that 
$$
[\langle{\pmb a},{\pmb u}\rangle, \langle{\pmb b},{\pmb u}\rangle]_*=
i\h \langle{\pmb a}J,{\pmb b}\rangle.
$$

If $H_*$ is a linear function, then this is given also as follows: 
By a direct calculation of intertwiner, we see that     
\begin{equation}
  \label{eq:intwin}
I_{_K}^{^{K'}}(e^{\frac{1}{i\h}\langle{\pmb a},{\pmb u}\rangle})
=e^{\frac{1}{4i\h}{\pmb a}(K'{-}K)\,{}^t\!{\pmb a}}
e^{\frac{1}{i\h}\langle{\pmb a},{\pmb u}\rangle}. 
\end{equation}
Hence, $\{e^{\frac{1}{4i\h}{\pmb a}K\,{}^t\!{\pmb a}}
e^{\frac{1}{i\h}\langle{\pmb a},{\pmb u}\rangle}; 
K\in{\mathcal S}_{\mathbb C}(2m)\}$ is a family of mutually
isomorphic one parameter groups {cf. \cite{OMMY3},\cite{OMMY4},\cite{OMMY5}}. 
We have denoted this element symbolically by 
$e_*^{\frac{1}{i\h}\langle{\pmb a},{\pmb u}\rangle}$. Namely we denote   
\begin{equation}
  \label{eq:tempexp}
:e_*^{\frac{1}{i\h}\langle{\pmb a},{\pmb u}\rangle}:_{_K}
=e^{\frac{1}{4i\h}\langle{\pmb a}K,{\pmb a}\rangle}
e^{\frac{1}{i\h}\langle{\pmb a},{\pmb u}\rangle}
=e^{\frac{1}{4i\h}\langle{\pmb a}K,{\pmb a}\rangle
{+}\frac{1}{i\h}\langle{\pmb a},{\pmb u}\rangle}.  
\end{equation}

The fact stated below is an example how a transcendental element
depends on the expression parameters: 
\begin{prop}\label{special}
If ${\rm{Im}}\langle{\pmb a}K,{\pmb a}\rangle$ is negative, then the 
formula of Fourier transform gives  
$$
\int_{-\infty}^{\infty}
:e_*^{t\frac{1}{i\h}\langle{\pmb a},{\pmb u}\rangle}:_{_K}dt\in 
{\mathcal E}_{2+}({\Bbb C}^{2m}).
$$  
In particular, both 
$$
{:}\langle{\pmb a},{\pmb u}\rangle{:}_{_{K+}}^{-1}
{=}\frac{1}{i\h}\int_{-\infty}^{0}
:e_*^{t\frac{1}{i\h}\langle{\pmb a},{\pmb u}\rangle}:_{_K}dt ,\quad 
{:}\langle{\pmb a},{\pmb u}\rangle{:}_{_{K-}}^{-1}
{=}-\frac{1}{i\h}\int_{0}^{\infty}
:e_*^{t\frac{1}{i\h}\langle{\pmb a},{\pmb u}\rangle}:_{_K}dt
$$ 
are $*$-inverses of ${:}\langle{\pmb a},{\pmb u}\rangle{:}_{_{K}}$
under the calculation of $*_{_K}$-product.
${:}\langle{\pmb a},{\pmb u}\rangle{:}_{_{K\pm}}^{-1}$
belongs to ${\mathcal E}_{2+}({\Bbb C}^{2m})$. 
Two different inverses of a single element breaks apparently the 
associativity.
\end{prop}

Furthermore, 
if ${\rm{Im}}\langle{\pmb a}K,{\pmb a}\rangle$ is negative, then
$\sum_{n=-\infty}^{\infty}{:}e_*^{n\frac{1}{i\h}\langle{\pmb a},{\pmb u}\rangle}{:}_{_K}$ 
converges to give Jacobi's theta function 
$\Theta_3(\frac{1}{i\h}\langle{\pmb a},{\pmb u}\rangle)$ in \cite{OMMY3}.

\medskip
\noindent
{\bf  Anti-automorphism  and  Hermite structure}

Note that generators of the algebra $(W_{2m}, *)$ are only abstract
symbols. It does not make sense to say that these are complex
variables or not. To make it clear we fix an hermitian structure 
by introducing an anti-automorphism 
$x\to{x}^{\iota}$, called the {\bf Hermitian conjugate}, by defining 
$$
{u}^{\iota}_k{=}u_k, \quad k=1{\sim}2m,\quad {i}^{\iota}=-i, \quad {\h}^{\iota}=\h.
$$
An element $H\in W_{2m}$ is called an {\it hermite element} when 
${H}^{\iota}{=}H$, but its $K$-ordered expression ${:}H_*{:}_{_K}$ does not
have the property $\overline{{:}H_*{:}_{_K}}{=}{:}H_*{:}_{_K}$ in general.

The next identity is useful.
\begin{prop}
If $e_*^{tH_*}$, $e_*^{t{H}^{\iota}_*}$ and $e_*^{s{H}^{\iota}_*}{*}e_*^{tH_*}$ are defined for 
$s, t\in\mathbb R$, then  
$e_*^{t{H}^{\iota}_*}{=}(e_*^{tH_*})^{\iota}$.  In particular, 
$e_*^{tu_i}$ is an hermite element for every $t\in\mathbb R$.
\end{prop}

\noindent
{\bf Proof}.\,\,Set $f_t=e_*^{tH_*}$, $g_t=e_*^{t{H}^{\iota}_*}$. Then 
 $\frac{d}{dt}f_{-t}{=}(-H_*){*}f_{-t}$. It follows 
$\frac{d}{dt}f_{-t}^{\iota}{=}f_{-t}^{\iota}{*}(-{H}^{\iota}_*)$. 
Thus, we have 
$$
\frac{d}{dt}f_{-t}^{\iota}{*}g_t{=}f_{-t}^{\iota}{*}(-{H}_*^{\iota}{+}{H}_*^{\iota}){*}g_t{=}0.
$$
It follows $e_*^{tH_*^{\iota}}{=}(e_*^{tH_*})^{\iota}$ as $f_{-t}^{\iota}{*}g_t{=}1$.\hfill$\Box$ 

If $H_*$ is hermitian and $F_t{=}e_*^{itH_*}$ exists, then $F_t$ is
unitary, since we have the primitive conservation law:
\begin{equation}\label{consevation}
\frac{d}{dt}{F^\iota}_t{*}F_t=0, \quad {F^\iota}_0{*}F_0=1. 
\end{equation}

\subsubsection{The case $\rho \in {\Cal H}_{2m}[\nu]$}
Suppose first that $\rho$ mentioned in \S\,\ref{prel01} satisfies $\rho{\in}{\Cal H}_{2m}[\nu]$, then 
$\rho$ must be a linear function 
$$
\langle{\pmb a},\tilde{\pmb
  u}\rangle{=}\sum_{i=1}^{m}(a_iu_i{+}a_{m{+}i}v_i).
$$
By Proposition\,\ref{special}, if ${\rm{Im}}\langle{\pmb a}K,{\pmb a}\rangle$ is negative,
then  $\langle{\pmb a},\tilde{\pmb u}\rangle$ is invertible. 

\bigskip 
Suppose $\rho{=}u_1$ for simplicity and denote its inverse by
$u_{1\ctt}^{-1}$. Then, the algebra generated by \eqref{nicecont} in 
\S\,\ref{prel01} turns out to be generated by 
\begin{equation}\label{newgenerator}
u_{1\ctt}^{-2}{*}\nu,\,\,u_{1\ctt}^{-1}{*}{v}_1,\,\, 
u_{1\ctt}^{-1}{*}{u}_2,\,\cdots,\,u_{1\ctt}^{-1}{*}{u}_m,\,\, 
u_{1\ctt}^{-1}{*}{v}_2,\,\cdots,\,u_{1\ctt}^{-1}{*}{v}_m
\end{equation}
where the calculations are executed by the $*_{_K}$-product formula under
the $K$-ordered expression. Thus, the suffix such as $*_{_K}$, ${:}\,\,{:}_{_K}$
are omitted.
 
For simplicity we denote these as follows:
$$
\mu,\,\,\tau,\,\,\tilde{u}_k,\,\,\tilde{v}_k, \,\, k{=}2{\sim}m 
$$
and let ${\Cal A}$ be the algebra generated by these in  
${\mathcal E}_{2+}({\Bbb C}^{2m})$.
Using $[u_{1\ctt}^{-1}, v_1]{=}i\h u_{1\ctt}^{-2}$, 
$[u_{1\ctt}^{-1}, u_k]{=}0{=}[u_{1\ctt}^{-1}, v_k]$, $k=2{\sim}m$, we
have 
$$
[\mu^{-1},\tau]{=}2i,\quad 
[\tilde{u}_i,\,\,\tilde{v}_j]{=}\mu i\delta_{ij},\quad 
[\tau, \tilde{u}_i]{=}\mu i\tilde{u}_i, \quad 
[\tau, \tilde{v}_i]{=}\mu i\tilde{v}_i,\quad {\text{all others are}}\,\, 0.
$$

As $[\mu,{\tilde u}_i]{=}0{=}[\mu,{\tilde v}_i]$, $i{=}2{\sim}m$,  
$\{\mu, {\tilde u}_i, {\tilde v}_i\}$ generates the Heisenberg algebra
${\Cal H}_{2m}[\mu]$. Hence 
${\Cal A}$ is a $\mu$-regulated algebra 
${\Cal H}_{2m}[\mu,\tau]$.
As $[\tau,\mu^{-1}]{=}-2i$, $\tau$ may be viewed 
as a canonical conjugate of $\mu^{-1}$.   

Now join $u_1$ to ${\Cal H}_{2m}[\mu,\tau]$. This is also a 
$\mu$-regulated algebra.

\begin{prop}\label{nongo} 
A suitably extended $\mu$-regulated algebra such as 
${\Cal H}_{2m}[\mu, u_1, \tau]$ contains no standard
abstract vacuum $($cf. \S\,\ref{sec:absvac}$)$.
\end{prop}

\noindent
{\bf Proof}.\,\,Suppose there is a standard abstract vacuum $\varpi$
in $\varpi\in {\Cal H}_{2m}[\mu, u_1, \tau]$. It follows 
$\varpi{*}u_1{=}0$ or $u_1{*}\varpi{=}0$. Hence we have 
$\frac{d}{ds}\varpi{*}e_*^{su_1}{*}\varpi{=}0$ and hence 
$\varpi{*}e_*^{su_1}{*}\varpi{=}\varpi$. 
Now recall the formula
$$
u_{1\ctt}^{-1}{*}u_{1\ctt}^{-1}{=}\int_{-\infty}^02s{:}e_*^{su_1}{:}_{_K}ds.
$$
Using this we have 
$$
\varpi{*}u_{1\ctt}^{-1}{*}u_{1\ctt}^{-1}{*}\varpi{=}
\varpi\,{*}\!\int_{-\infty}^{0}2se_*^{su_1}{*}\varpi ds{=}
\int_{-\infty}^{0}2s\varpi{*}e_*^{su_1}{*}\varpi ds{=}\infty\varpi.
$$
It follows $\varpi{*}\mu{*}\varpi{=}\infty\varpi$. \hfill $\Box$

\medskip
Apparently, this is caused by the choice 
$u_{1\ctt}^{-1}{\in}{\mathcal E}_{2+}({\Bbb C}^{2m})$.   
If this is replaced by the embedded Heisenberg algebra given in  
\S\,\ref{embedding}, then we can find an abstract vacuum in it
(see \S\,\ref{embedded}).

\subsubsection{Heisenberg algebra embedded in Weyl algebra}\label{embedding}
In what follows we denote by   
\begin{equation}\label{coordinate} 
x_0, x_1,\cdots, x_m, y_0, y_1, \cdots, y_{m}
\end{equation} 
natural coordinate functions of ${\mathbb C}^{2m{+}2}$. 
Consider the Weyl algebra  $(W_{2m{+}2}[\h],*)$ generated by setting  
the relation $[x_i,y_j]=-i\h\delta_{ij}$. 
Fix a generic expression parameter
$K\in {\mathcal S}_{\mathbb C}(2m{+}2)$. In what follows, all
calculations executes under the $K$-ordered expression and
$*_{_K}$-product formula. We first embed the Heisenberg algebra 
${\Cal H}_{2m}[\nu]$ into  
$({\mathcal E}_{1+}({\Bbb C}^{2m{+}2}, {*_{_K}})$.

\medskip
In this algebra, we make first the $*$-exponential function
$e_*^{ty_0}$. As it was seen in \S\,\ref{starexp00}, the exponential law
$e_*^{sy_0}{*}e_*^{ty_0}{=}e_*^{(s{+}t)y_0}$ holds and the $K$-ordered
expression is 
${:}e_*^{ty_0}{:}_{_K}{=}e^{\frac{i\h}{4}\delta'}e^{e^{ty_0}}$ by
using a suitable constant $\delta'$.  
Hence $e_*^{sy_0}$ is an element of ${\mathcal E}_{1+}({\Bbb C}^{2m{+}2})$.

\medskip
Set $\mu=\h{*}e_*^{-2y_0}$. As $\h{>}0$, the exponential law and
Theorem\,\ref{main01} show that  
$\h e_*^{-2y_0}*: {\mathcal E}_{1+}({\Bbb C}^{2m{+}2})
\to {\mathcal E}_{1+}({\Bbb C}^{2m{+}2})$ is a linear isomorphism. 

We set furthermore 
\begin{equation}\label{notation}
\mu=\h e_*^{-2y_0}, \quad 
\tau=\frac{1}{2}(e_*^{-2y_0}{*}x_0{+}x_0{*}e_*^{-2y_0}), 
\quad {\tilde u}_i=e_*^{-y_0}{*}x_i,\quad {\tilde v}_i=e_*^{-y_0}{*}y_i,\quad i=1\sim m.
\end{equation}
They are all hermite elements (cf.\S\,\ref{starexp00}) and 
$$
\tau=e_*^{-2y_0}{*}(x_0{+}i\h){=}(x_0{-}i\h){*}e_*^{-2y_0},
$$
as $[e_*^{-2y_0},x_0]=-2i\h e_*^{-2y_0}{=}-2i\mu$. Using these we have 
\begin{equation}\label{tubularnbh}
[\tau, \mu]=2i\mu_*^2,\quad [\mu,{\tilde u}_i]{=}0,\quad  
[\mu,\tilde{v}_i]{=}0,\quad  
[\tau,\tilde{u}_i]=\mu i{*}\tilde{u}_i, \quad 
[\tau,\tilde{v}_i]=\mu i{*}\tilde{v}_i,\quad 
[\tilde{u}_i,\tilde{v}_j]=-i\mu\delta_{ij}.
\end{equation}
As $[\mu{*}\mu^{-1},\tau]=0$, we see $[\mu^{-1},\tau]=2i$. 
The algebra generated by these will be denoted by 
$\widetilde{\Cal H}_{2m}[\mu,\tau]$.  
It is clear that 
$\widetilde{\Cal H}_{2m}[\mu,\tau]$ contains the Heisenberg algebra. 
Since $\mu^{-1}{\ctt}\tau=\frac{1}{\h}x_1$, $E_t$ is defined as the automorphism 
${\rm{Ad}}(e_*^{t\frac{1}{i\h}x_1})$. 

\bigskip
On the other hand, note that   
 $[\h^{-1}e_*^{2y_0}, e_*^{-2y_0}{*}x_0]{=}2i$, 
although $e_*^{-2y_0}{*}x_0$ is not hermitian. Hence 
$\hat\tau{=}e_*^{-2y_0}{*}x_0$ may be regarded as a canonical
conjugate of $\mu^{-1}$. 
Since $\mu^{-1}{*}\hat\tau=\frac{1}{\h}x_0$, we
see ${\rm{Ad}}(e_*^{t\frac{1}{i\h}x_0}){=}E_t$.
Thus, by using $\hat\tau$
instead of $\tau$, we have the same commutation relations as 
\eqref{tubularnbh}.

\medskip
Summarizing these we see the next 
\begin{thm}\label{thm}
Heisenberg algebra ${\Cal H}_{2m}[\mu]$ is isomorphic to the
subalgebra of $\widetilde{\Cal H}_{2m}[\mu,\tau]\,
(\subset 
{\mathcal E}_{1+}({\Bbb C}^{2m{+}2}))$ 
where $\mu$ corresponds to $\h e_*^{-2y_0}$,
and $\frac{1}{2}(e_*^{-2y_0}{*}x_0{+}x_0{*}e_*^{-2y_0})$ is a
canonical conjugate of $\mu_*^{-1}$. 

\medskip
On the other hand, ${\Cal H}_{2m}[\mu]$ is also isomorphic to the
subalgebra of $\widetilde{\Cal H}_{2m}[\mu,\hat\tau]$, although 
$\hat\tau$ is not an Hermite element.

\medskip
Besides the topological completeness, 
$\widetilde{\Cal H}_{2m}[\mu,\tau]$ $($resp.   
$\widetilde{\Cal H}_{2m}[\mu,\hat\tau]$$)$
satisfies {\rm{(A.1)}}$\sim${\rm{(A.4)}}
and all properties of contact Weyl algebra, 
where $\frac{1}{2}\partial_{\tau}$
$($resp.$\frac{1}{2}\partial_{\hat\tau}$$)$
is the characteristic vector field 
and 
$B{=}{\Cal H}_{2m}[\tau]$ $($resp.${\Cal H}_{2m}[\hat\tau]$$)$, 
\end{thm}

In \S\,\ref{embedded}, a vacuum representation of 
$\widetilde{\Cal H}_{2m}[\mu,\tau]$ will be given. 
It is rather surprising that ${:}e_*^{it\tau}{:}_0$ is 
defined by a difference-differential equation,  and  
${\rm{Ad}}(e_*^{it\tau})$ is defined only for $t{\geq}0$. 
Such a semi-group property is familiar in heat equation. 
But, this appears clearer in the calculation 
of $e_*^{it\tau}$ under the Moyal product formula.

\bigskip

\subsection{$*$-exponentials of quadratic forms}

To define $\sqrt[*]{\rho_*^2}$ mentioned in \S\,\ref{prel01}, 
we have to use the formula of Laplace transform 
$$ 
\sqrt[*]{\rho_*^2}^{-1}{=}\frac{1}{\sqrt{\pi}}\int_{0}^{\infty}\frac{1}{\sqrt{t}}e_*^{-t\rho_*^2}dt.
$$
Thus, we have to know first the formula of ${*}$-exponential functions
of quadratic forms.

As $(W_{2m{+}2}[\h],{*})=(W_2[\h],{*}){\otimes}(W_{2m}[\h],{*})$, main properties
of Weyl algebra are explained in $(W_2[\h],{*})$. 
Let $(W_2[\h],{*})$ be the Weyl algebra generated by $x, y$. 

\medskip
Let 
$K{=}
\footnotesize{
\begin{bmatrix}
\delta&c\\
c&\delta'
\end{bmatrix}}
$ 
be the expression parameter. In this section, we summarize properties of 
some of $*$-exponential functions of quadratic forms. 
The $K$-ordered expression of the $*$-exponential function $e_*^{t\frac{1}{i\h}2x{\ctt}y}$
is given as follows: 
By setting $\Delta{=}e^t{+}e^{-t}{-}c(e^t{-}e^{-t})$, it is given by   
\begin{equation}\label{genericparam00}
{:}e_*^{t\frac{1}{i\h}2x{\ctt}y}{:}_{_{K}}{=}
\frac{2}{\sqrt{\Delta^2{-}(e^t{-}e^{-t})^2\delta\delta'}} 
\,\,e^{\frac{1}{i\h}
\frac{e^t-e^{-t}}{\Delta^2{-}(e^t{-}e^{-t})^2\delta\delta'}
\big((e^t-e^{-t})(\delta'{x}^2{+}\delta{y}^2){+}2\Delta xy\big)}.
\end{equation}
Setting $\delta\delta'=\rho^2$, we see 
\begin{equation}\label{twolines}
\sqrt{\Delta^2{-}(e^t{-}e^{-t})^2\delta\delta'} =
e^{-t}
\sqrt{((1{-}c{+}\rho)e^{2t}{+}(1{+}c{-}\rho)) 
((1{-}c{-}\rho)e^{2t}{+}(1{+}c{+}\rho))}.
\end{equation}

Set $H_*{=}\frac{1}{i\h}2x{\ctt}y$. 
By \eqref{genericparam00}, we easily see  the following 
under generic expressions:

\bigskip
\noindent
({\bf a})\,\,$e_*^{tH_*}$ is rapidly decreasing on $\mathbb R$ of
$e^{-|t|}$ order. 

\bigskip
\noindent
({\bf b})\,\,$e_*^{zH_*}$ is $4\pi i$-periodic 
${:}e_*^{(z{+}4\pi i)H_*}{:}_{_K}={:}e_*^{zH_*}{:}_{_K}$. In precise 
for generic $K$-expressions there are $a, b$ 

depending on $K$ such that $-\infty{<}a{<}b{<}\infty$ and 

\medskip
({\bf b,1}) if $a<s<b$, then  
 ${:}e_*^{(s{+}it)H_*}{:}_{_K}$ is $2\pi$-periodic.  

\medskip
({\bf b,2}) if $s{<}a$ or $b{<}s$, then
${:}e_*^{(s{+}it)H_*}{:}_{_K}$ is alternating $2\pi$-periodic.

\bigskip
\noindent
({\bf c}) $\lim_{t\to{-}\infty}{:}e^{-t}e_*^{tH_*}{:}_{_K}$ is a nontrivial
 element denoted by ${:}{\varpi}_{00}{:}_{_K}$. We call this the 
 {\bf vacuum}. 

\bigskip
In precise,  the path tending to  $-\infty$ should be in the 
same sheet as in the origin $1$, as  $*$-exponential functions 
 have double branched singularities as it is seen in
 \eqref{genericparam00}.
By the exponential law $e^{t}e_*^{tH_*}{=}e_*^{t(H_*{+}1)}$, ${\varpi}_{00}$ is idempotent, i.e. 
${\varpi}_{00}{*}{\varpi}_{00}{=}{\varpi}_{00}$   
and the bumping identity $y{*}(x{\ctt}y){=}(x{\ctt}y{+}i\h){*}y$ give the property that 
$$
y{*}{\varpi}_{00}{=}0{=}{\varpi}_{00}{*}x.
$$  
Similarly, a nontrivial element 
$\overline{\varpi}=\lim_{t\to{}\infty}e^{t}e_*^{tH_*}$
is called the {\bf bar-vacuum}. 
This is an idempotent element such that 
${x}{*}\overline{\varpi}_{00}{=}0{=}\overline{\varpi}_{00}{*}y.$

\bigskip
\noindent
({\bf d}) Let $w{=}{x}{+}i{y}$, 
$\overline{w}{=}{x}{-}i{y}$. Then
$w{\ctt}\overline{w}{=}x_*^2{+}y_*^2$, 
$[w, \overline{w}]{=}-2\h$. By a similar 
calculation gives (cf. \cite{OMMY5})
\begin{equation}\label{Delta}
{:}e_*^{t\frac{1}{\h}(x_*^2{+}y_*^2)}{:}_{_K}{=}
\frac{1}{\sqrt{\Delta_{K}(t)}}
e^{\frac{1}{i\h}\frac{i\sinh t}{\Delta_{K}(t)}
\big((\cosh t{-}\delta'i\sinh t)u^2{+}(\cosh t{-}\delta i\sinh t)v^2
{+}2ci(\sinh z) uv\big)}
\end{equation}
where 
$\Delta_{K}(t){=} 
\cosh^2t{-}(\delta{+}\delta')i\sinh t\cosh t{+}(c^2{-}\delta\delta')\sinh^2t.$

\medskip
In generic ordered expression 
the {\bf complex vacuum} ${:}{\varpi}_{\mathbb C}{:}_{_K}$ is defined by 
\begin{equation}\label{cmpvac}
{:}{\varpi}_{\mathbb C}{:}_{_K}{=}
\lim_{t\to{-}\infty}{:}e^{-t}e_*^{t\frac{1}{\h}w{\ctt}\overline{w}}{:}_{_K}. 
\end{equation}
This exists by \eqref{Delta} and  this  is an idempotent element satisfying 
$\overline{w}{*}{\varpi}_{\mathbb C}{=}0{=}{\varpi}_{\mathbb C}{*}w$, 
providing the path tending to $-\infty$ avoiding singular points belongs to the 
same sheet as in the origin $1$.

\medskip
In spite of the double valued nature caused by $\sqrt{\,\,}$, the
exponential law holds by computations such as 
$\sqrt{\alpha}\sqrt{\beta}{=}\sqrt{\alpha\beta}$.
If $K{=}0$, then it is called the {\bf Weyl ordered expression}, and if $c{=}1, 
\delta{=}0{=}\delta'$, then it is called the {\bf normal ordered expression}.
These are often used in physics, but general $K$-expressions are not
used in physics.

\medskip
For a degenerate quadratic form, the formula of the $*$-exponential
functions are given by the case of one variable, and hence  
$$
{:}e_*^{t\frac{1}{i\h}x_*^2}{:}_{_K}{=}
\frac{\sqrt{i\h}}{\sqrt{i\h{-}t\delta}}e^{\frac{i\h t}{i\h{-}t\delta}x^2}.
$$ 
Similar to vacuums, there exists the limit   
$$
\lim_{t\to\infty}{:}\sqrt{t}e_*^{t\frac{1}{i\h}x_*^2}{:}_{_K}{=}
\frac{\sqrt{i\h}}{\sqrt{{-}\delta}}e^{-\frac{i\h}{\delta}x^2}
$$ 
and 
${:}e_*^{t\frac{1}{i\h}x_*^2}{:}_{_K}{*_{_K}}
\frac{\sqrt{i\h}}{\sqrt{{-}\delta}}e^{-\frac{i\h}{\delta}x^2}
{=}
\frac{\sqrt{i\h}}{\sqrt{{-}\delta}}e^{-\frac{i\h}{\delta}x^2}$.
However, as $\sqrt{t}e_*^{t\frac{1}{i\h}x_*^2}$ is not a one parameter
subgroup, the limit is not idempotent, but  
$\lim_{t\to\infty}(\sqrt{t}e_*^{t\frac{1}{i\h}x_*^2})_*^2$ diverges.

\subsubsection{Abstract  vacuums on several variables}\label{comment2}
Let $x_0, x_1,\cdots, x_m, y_0, y_1,\cdots, y_m$ be generators of $W_{2m{+}2}[\h]$.
We want to apply the formula for the case $m{=}1$ mentioned above to the $*$-exponential functions 
of quadratic forms of $2m$-variables.  

Obviously, the $K$-expression 
${:}f_*(x_i,y_i){*}g(x_j,y_j){:}_{_K}$ of $f_*(x_i,y_i){*}g(x_j,y_j)$
is computed by using only submatrices relating $(i,j)$ components. For
a given $K$, let 
$$
K_{(i)}{=}\footnotesize{
\begin{bmatrix}
K_{i,i}&K_{i,m{+}i}\\
K_{m{+}i,i}&K_{m{+}1,m{+}i}
\end{bmatrix}
}, \quad 
K_{(i,j)}{=}\footnotesize{
\begin{bmatrix}
K_{i,j}&K_{i,m{+}j}\\
K_{m{+}j,i}&K_{m{+}i,m{+}j}
\end{bmatrix}
}.
$$
Then it is clear that 
$$
{:}f_*(x_i,y_i){*}g_*(x_j,y_j){:}_{_K}{=}
{:}f_*(x_i,y_i){:_{_{K_{(i)}}}}{*_{_{K_{(i,j)}}}}{:}g_*(x_j,y_j){:}_{_{K_{(j)}}}.
$$
Recall now how the product $e_*^{H_*}{*}F$ is defined. This is defined
by the value at $t=1$ of the real analytic solution of  
$$
\frac{d}{dt}f_t=H_*{*}f_t, \quad f_0{=}F.
$$
$e_*^{H_{*k}}{*}\cdots{*}e_*^{H_{*2}}{*}e_*^{H_{*1}}{*}F$
are defined similarly. Such products may be called 
the {\it path connecting products}. Hence in general  
$e_*^{H_{*2}}{*}e_*^{H_{*1}}{*}1$ might be different from
$e_*^{H_{*1}}{*}e_*^{H_{*2}}{*}1$, even if $[H_{*1}, H_{*2}]_*{=}0$.
Recalling \S\,\ref{comments}, the next Proposition is not trivial.
\begin{prop}\label{comment}
In generic ordered expression, we have 
$$
e_*^{ta_1x_1{\ctt}y_1}{*}e_*^{ta_2x_2{\ctt}y_2}{*}\cdots{*}e_*^{ta_mx_m{\ctt}y_m}
{=}
e_*^{t(a_1x_1{\ctt}y_1{+}a_2x_2{\ctt}y_2{+}\cdots{+}a_mx_m{\ctt}y_m)}
$$
if each component has no singular point on the interval $[0,t]$.
\end{prop}

\medskip

Recall the remark given in \S\,\ref{comment2} about how to 
apply the formulas in the case $m=1$ to
the case $m\geq 2$. 

In what follows, several abstract vacuums which can be joined to the 
algebra ${\mathcal E}_{1+}({\Bbb C}^{2m{+}2})$ without trouble,
although they are in ${\mathcal E}_{2+}({\Bbb C}^{2m{+}2})$.

Let 
\begin{equation}\label{standardvac}
\widetilde{\varpi}(L_0){=}\varpi_{00}{*}{\widetilde{\varpi}(L)}{=}
\varpi_{00}{*}\prod_{k=1}^m{*}\varpi_{00}(k),\quad 
\widetilde{\varpi}(\overline{L}_0){=}\overline{\varpi}_{00}{*}{\widetilde{\varpi}(L)}{=}
\overline{\varpi}_{00}{*}\prod_{k=1}^m{*}\varpi_{00}(k),
\end{equation}
where
$\varpi_{00}(k){=}\lim_{t\to-\infty}e_*^{t\frac{1}{i\h}x_k{\ctt}y_k}$,
\quad 
$\overline{\varpi}_{00}{=}\lim_{t\to\infty}e^{t\frac{1}{2}}e_*^{t\frac{1}{i\h}x_0{\ctt}y_0}$.
These have properties as follows:
$$
y_k{*}\widetilde{\varpi}(L_0){=}0, \,\,(k{=}0{\sim}m), \quad 
x_0{*}\widetilde{\varpi}(\overline{L}_0){=}0{=}\widetilde{\varpi}(\overline{L}_0){*}y_0,\,\,
y_k{*}\widetilde{\varpi}(\overline{L}_0){=}0, \,\,(k{=}1{\sim}m).
$$
Similarly, on ${\mathbb C}^m$, the {\bf complex vacuum}
$\widetilde{\varpi}_{\mathbb C}$ is given by 
\begin{equation}\label{cplxvac}
\widetilde{\varpi}_{\mathbb C}{=}
\lim_{s\to{-}\infty}e_*^{s\sum\zeta_i{*}\overline{\zeta}_i}. 
\end{equation}
This satisfies 
$\overline{\zeta}_i{*}\widetilde{\varpi}_{\mathbb C}{=}0$. 

\medskip
As it was mentioned in \S\,\ref{prel01}, there are various abstract
vacuums. As 
$\widetilde{\varpi}(\overline{L}_0){*}e_*^{y_0}{*}\widetilde{\varpi}(\overline{L}_0)
{=}\widetilde{\varpi}(\overline{L}_0)$, we see 
$e_*^{y_0}{*}\widetilde{\varpi}(\overline{L}_0)$ is an abstract vacuum
such that by using $(x_0{+}i\h)e_*^{y_0}{=}e_*^{y_0}{*}x_0$  
$$
(x_0{+}i\h)e_*^{y_0}{*}\widetilde{\varpi}(\overline{L}_0){=}0{=}
\tau{*}e_*^{y_0}{*}\widetilde{\varpi}(\overline{L}_0).
$$ 
We denote this by 
\begin{equation}\label{contvac}
\widetilde{\varpi}_{y_0}(\overline{L}_0){=}e_*^{y_0}{*}\widetilde{\varpi}(\overline{L}_0)
\end{equation}

In what follows, we use the same notations as in \eqref{notation}. 
Under the notation \eqref{notation}, $e_*^{is\tau}{*}\varpi_{00}$ has 
the property as follows:
\begin{prop}
 For $|2s\h|{<}1$, 
$\varpi_{00}{*}e_*^{-is\tau}{*}\varpi_{00}{=}\frac{1}{\sqrt{1{+}2s\h}}\varpi_{00}$. 
In particular, $\sqrt{1{+}2s\h}\,e_*^{is\tau}{*}\varpi_{00}$ is an 
idempotent element such that 
$$
\varpi_{00}{*}x_0{*}\sqrt{1{+}2s\h}e_*^{is\tau}{*}\varpi_{00}{=}0,\quad 
\varpi_{00}{*}y_0{*}\sqrt{1{+}2s\h}e_*^{is\tau}{*}\varpi_{00}
{=}i\h\sqrt{1{+}2s\h}e_*^{is\tau}{*}\varpi_{00}.
$$ 
In particular, $e_*^{-is\tau}{*}\widetilde{\varpi}(L_0)$ is an
abstract vacuum for $|2s\h|{<}1$.
\end{prop}

\noindent
{\bf Proof}\,\,Note that $\varpi_{00}{*}x_0=0$, $e_*^{-2y_0}{*}{\varpi}_{00}{=}{\varpi}_{00}$.
Computing $\varpi_{00}{*}(\tau_*^n){*}\varpi_{00}$ gives    
$$
\varpi_{00}{*}(-is\tau_*)^n{*}\varpi_{00}{=}(\frac{1}{2})_n(-s\h)^n{*}\varpi_{00},\quad
(a)_n=a(a{+}1)\cdots(a{+}n{-}1).
$$
Hence 
$$
\sum_n\varpi_{00}{*}\frac{1}{n!}(-is\tau_*)^n{*}\varpi_{00}{=}
\sum_n\frac{1}{n!}(\frac{1}{2})_n(-s\h)^n\varpi_{00}{=}\frac{1}{\sqrt{1{+}2s\h}}{*}\varpi_{00},\quad
|2s\h|<1.
$$  
In \eqref{justext} later, the condition $|2s\h|{<}1$ is relaxed to
$1{+}2s\h >0$. ${  }$\hfill $\Box$ 

\bigskip
We denote this by 
\begin{equation}\label{Heisenvac}
\widetilde{\varpi}_{s}(\Cal H){=}e_*^{-is\tau}{*}\widetilde{\varpi}(L_0).
\end{equation}

$\widetilde{\varpi}_{y_0}(\overline{L}_0)$ and 
$\widetilde{\varpi}_{s}(\Cal H)$ 
will be used in \S\,\ref{embedded}.

\subsection{Reductions vs  Restriction  to  the Energy surface}\label{RedRes}
Consider the most familiar case $\rho_*^2{=}\sum_{k=1}^m(u_k{*}u_k{+}v_k{*}v_k)$
mentioned in \S\,\ref{prel01}. 
Set 
$$
\zeta_k{=}u_k{+}iv_k,\quad  
\bar\zeta_k{=}u_k{-}iv_k, \quad k=1{\sim}m.
$$
Then, $\rho_*^2{=}\sum_{k=1}^m\zeta_k{\ctt}\bar\zeta_k$, where $a{\ctt}b{=}\frac{1}{2}(a{*}b{+}b{*}a)$.
First of all, note that \eqref{Delta} and the comment given in \S\,\ref{comment2} give the following:

\begin{prop}
In generic ordered expressions 
$e_*^{t\rho_*^2}{=}\prod_{k=1}^m{*}e_*^{t\zeta_k{\ctt}\bar\zeta_k}$
and 
$\int_{0}^{\infty}e_*^{-t\rho_*^2}dt$ is defined to give
an inverse of $\rho_*^2$. Moreover, the Laplace transform gives 
$$
\rho_*^{-1}{=}
\sqrt[*]{\rho_*^2}^{-1}{=}
\frac{1}{\sqrt{\pi}}\int_{0}^{\infty}\frac{1}{\sqrt{t}}e_*^{-t\rho_*^2}dt
$$ 
\end{prop}
As $\rho_*^2$ has the property $E_s(\rho_*^2){=}e^{2s}\rho_*^2$, 
changing variables in the integration gives easily 
$E_s\rho_*^{-1}{=}e^{-s}\rho_*^{-1}$.

\bigskip
$E_s$-invariant elements are generated by 
$$
\nu{*}\rho_*^{-2},\quad \rho_*^{-1}{*}\zeta_k, \quad
\bar\zeta_k{*}\rho_*^{-1},\quad k=1{\sim} m. 
$$
We denote the generated algebra by ${\Cal A}(S^{2m-1})$ in the space
${\mathcal E}_{2+}({\mathbb C}^{2m})$. 

For simplicity we denote 
$$
\mu{=}\nu{*}\rho_*^{-2},\quad \xi_k{=}\rho_*^{-1}{*}\zeta_k,\quad 
\bar\xi_k{=}\bar\zeta_k{*}\rho_*^{-1},\quad k=1{\sim}m.
$$
Although they are not hermitian, we see easily 
\begin{equation}\label{heisenmu}
[\mu^{-1}, \xi_k]{=}\xi_k,\quad 
[\mu^{-1}, \bar\xi_k]{=}{-}\bar\xi_k, \quad 
\sum_{k=1}^m\xi_k{*}\bar\xi_k{=}1{-}2m\mu.
\end{equation}
It follows that $e^{it{\rm{ad}}(\mu^{-1})}(\xi_k){=}e^{it}\xi_k$,
$e^{it{\rm{ad}}(\mu^{-1})}(\bar\xi_k){=}e^{-it}\bar\xi_k$, and  
${\rm{Ad}}(e_*^{it\mu^{-1}})$ causes an $S^1$-action on 
${\Cal A}(S^{2m-1})$.
Hence $S^1$-invariant elements form a subalgebra 
${\Cal A}(P_{m-1}({\mathbb C}))$ generated by 
$$
\mu,\quad \xi_k{*}\bar\xi_{\ell},\quad \bar\xi_{k}{*}\xi_{\ell},\quad
1\leq k,\ell\leq m.   
$$

\bigskip

Furthermore, as $\rho_*^2{*}\zeta_k{=}\zeta_k{*}(\rho_*^2{+}2\nu)$, 
$\rho_*^2{*}\bar\zeta_k{=}\zeta_k{*}(\rho_*^2{-}2\nu)$,  we have 
$$
\rho_*^{-1}{*}\zeta_k{=}
\frac{1}{\sqrt{\pi}}\int_{0}^{\infty}\frac{1}{\sqrt{t}}e_*^{-t\rho_*^2}{*}\zeta_kdt{=}
\zeta_k{*}\frac{1}{\sqrt{\pi}}\int_{0}^{\infty}\frac{1}{\sqrt{t}}e_*^{-t(\rho_*^2{+}2\nu)}dt
{=}\zeta_k{*}\frac{1}{\sqrt{\rho_*^2{+}2\nu}}.
$$
Using $\sqrt{\rho_*^2{+}2\nu}{=}\rho_*{*}\sqrt{1{+}2\mu}{=}\sqrt{1{+}2\mu}{*}\rho_*$, we have 
$$
\rho_*^{-1}{*}\zeta_k{=}\zeta_k{*}\rho_*^{-1}{*}\sqrt{1{+}2\mu}^{-1}
$$
Similar calculation gives 
$$
\bar\zeta_k{*}\rho_*^{-1}{=}\sqrt{1{+}2\mu}^{-1}{*}\rho_*^{-1}{*}\bar\zeta_k,
\quad 
\rho_*^{-1}{*}\bar\zeta_k{=}\bar\zeta_k{*}\rho_*^{-1}{*}\sqrt{1{-}2\mu}^{-1},
\quad 
\zeta_k{*}\rho_*^{-1}{=}\sqrt{1{-}2\mu}^{-1}{*}\rho_*^{-1}{*}\zeta_k.
$$
By using first two identities (without using $\sqrt{1{-}2\mu}$) we
have that 
\begin{equation}\label{comm}
[\xi_k,\xi_{\ell}]{=}0{=}[\bar\xi_k,\bar\xi_{\ell}], \quad 
[\xi_k,\bar\xi_{\ell}]{=}-\frac{1}{1{+}2\mu}{*}(2\mu\delta_{k\ell}{+}\xi_k{*}\bar\xi_{\ell})
\end{equation} 
where we used the fact that 
$$
[\mu^{-1}, \zeta_k{*}\bar\zeta_{\ell}]{=}\nu^{-1}{*}[\rho_*^2, \zeta_k{*}\bar\zeta_{\ell}]{=}0,
\quad [\sqrt{1{+}2\mu}, \rho]{=}0{=}[\sqrt{1{+}2\mu}, \zeta_k{*}\bar\zeta_{\ell}].
$$

\begin{prop}\label{mureg}
The algebra ${\Cal A}(S^{2m{-}1})$ generated by $\{\mu, \,\xi_k,\, \bar\xi_k; k=1{\sim}m\}$
is a contact Weyl algebra such that 
${\rm{ad}}(i\mu^{-1})$ gives an $S^1$-action on ${\Cal A}(S^{2m{-}1})$.     
Commutation relations are given by \eqref{heisenmu} and \eqref{comm}. 
\end{prop}

For every non-negative integer $n{\in}{\mathbb N}$, we see 
$[\mu^{-1}, \xi_{i*}^n]{=}n\xi_{i*}^n$, $[\mu^{-1}, \bar\xi_{i*}^n]{=}-n\bar\xi_{i*}^n$.  
Hence ${\Cal A}(S^{2m{-}1})$ has the eigenspace decomposition of ${\rm{ad}}(\mu^{-1})$
$$
{\Cal A}(S^{2m{-}1}){=}\sum_{\ell\in\mathbb Z}\oplus{\Cal A}_{\ell}(S^{2m{-}1}), \quad 
{\Cal A}_{\ell}(S^{2m{-}1}){*}{\Cal A}_{\ell'}(S^{2m{-}1}){=}{\Cal A}_{\ell+\ell'}(S^{2m{-}1}),
\quad \ell, \ell'\in\mathbb Z.
$$
The subalgebra 
${\Cal A}_{0}(S^{2m{-}1})={\Cal A}(P_{m{-}1}(\mathbb C))$ 
 may be viewed as the algebra obtained by the reduction procedure. 

\subsubsection{Complex vacuum representation}

Computations in this subsection execute in generic ordered
expressions and the suffix ${:}\,\,{:}_{_K}$ will be omitted.
Here we use the vacuum \eqref{cplxvac}
Hence we have 
$\overline{\xi}_i{*}\widetilde{\varpi}_{\mathbb C}{=}0,\,\,
\sum_{k=1}^m \xi_k{*}\bar\xi_k{*}\widetilde{\varpi}_{\mathbb C}{=}
(1{-}2m\mu){*}\widetilde{\varpi}_{\mathbb C}{=}0$.
 
As $\rho_*^{2}{*}\widetilde{\varpi}_{\mathbb  C}
{=}2m\nu\widetilde{\varpi}_{\mathbb C}$, we have 
$$
\rho_*^{-1}{*}\widetilde{\varpi}_{\mathbb C}{=}
\frac{1}{\sqrt{\pi}}\int_{0}^{\infty}
\frac{1}{\sqrt{t}}e^{-t2m\nu}dt\widetilde{\varpi}_{\mathbb C}{=}
\frac{1}{\sqrt{2m\nu}}\widetilde{\varpi}_{\mathbb C},\quad
\sqrt{\mu}{*}\widetilde{\varpi}_{\mathbb C}{=}
\frac{1}{\sqrt{2m}}\widetilde{\varpi}_{\mathbb C}.
$$
$$
\rho_*^2{*}h(\pmb\zeta){*}\widetilde{\varpi}_{\mathbb C}
{=}\sum_k(\nu\zeta_k\partial_{k}{+}2\nu)h(\pmb\zeta){*}\widetilde{\varpi}_{\mathbb C}, \quad 
\mu^{-1}{*}h(\pmb\zeta){*}\widetilde{\varpi}_{\mathbb C}
{=}\sum_k(\zeta_k\partial_{k}{+}2)h(\pmb\zeta){*}\widetilde{\varpi}_{\mathbb C}.
$$
In particular, for every polynomial $p_k(\pmb\zeta)$ of degree $k$
$$
\mu^{-1}{*}p_k(\pmb\zeta){*}\widetilde{\varpi}_{\mathbb C}{=}
(k{+}2m)p_k(\pmb\zeta){*}\widetilde{\varpi}_{\mathbb C}, \quad 
\mu{*}p_k(\pmb\zeta){*}\widetilde{\varpi}_{\mathbb C}{=}
\frac{1}{k{+}2m}p_k(\pmb\zeta){*}\widetilde{\varpi}_{\mathbb C}
.$$
As 
$e_*^{s\sum\zeta_i{\ctt}\bar\zeta_i}
{*}p_k(\pmb\zeta){*}\widetilde{\varpi}_{\mathbb C}
{=}e^{s(k{+}2m\nu)}p_k(\pmb\zeta){*}\widetilde{\varpi}_{\mathbb C}$,
we have 
$$
\xi_i{*}p_k(\pmb\zeta){*}\widetilde{\varpi}_{\mathbb C}
{=}\frac{1}{\sqrt{k{+}1{+}2m\nu}}\zeta_ip_k(\pmb\zeta){*}\widetilde{\varpi}_{\mathbb C},\quad
\bar\xi_i{*}p_k(\pmb\zeta){*}\widetilde{\varpi}_{\mathbb C}
{=}\frac{1}{\sqrt{k{+}2m\nu}}\partial_ip_k(\pmb\zeta){*}\widetilde{\varpi}_{\mathbb C}.
$$
Hence, there is no canonical conjugate of $\mu^{-1}$.

\bigskip
One can use ${\mathbb C}[\pmb\xi]{*}\widetilde{\varpi}_{\mathbb C}$ as
the representation space. Then 
$$
\mu^{-1}{*}p_k(\pmb\xi){*}\widetilde{\varpi}_{\mathbb C}{=}
(k{+}2m)p_k(\pmb\xi){*}\widetilde{\varpi}_{\mathbb C}, \quad 
$$ 
The formula for 
$\bar\xi_i{*}p_k(\pmb\xi){*}\widetilde{\varpi}_{\mathbb C}$ is
complicated. 

\subsubsection{Localizations joining divisors} 
Consider now localizations by joining divisors, say 
$$
(\bar\zeta_k{*}\zeta_k)_{+}^{-1}, \quad k=1{\sim}m,
$$
where
$(\bar\zeta_k{*}\zeta_k)_{+}^{-1}{=}
\int_0^{\infty}e_*^{-t\bar\zeta_k{*}\zeta_k}dt$. 
$\bar\zeta_k^{\btt}{=}\zeta_k{*}(\bar\zeta_k{*}\zeta_k)_{+}^{-1}$ is a
left-inverse of $\zeta$ i.e. $\bar\zeta{*}\bar\zeta_k^{\btt}{=}1$, but 
$$
\bar\zeta_k^{\btt}{*}\bar\zeta{=}1{-}\lim_{t\to-\infty}e_*^{-t\zeta_k{*}\bar\zeta_k}
\quad {\text{(partial complex vacuum)}}.
$$

Recalling the generator system \eqref{newgenerator}, we consider 
a new generator system for $E_t$-invariant elements in generic ordered
expressions 
$$
\nu{*}(\zeta_1{*}(\bar\zeta_1{*}\zeta_1)_{+}^{-1})^2,\,\,
\frac{1}{2}\zeta_1^2{*}(\bar\zeta_1{*}\zeta_1)_{+}^{-1},\,\,
\,\,\zeta_k{*}\zeta_1{*}(\bar\zeta_1{*}\zeta_1)_{+}^{-1},\quad 
 \zeta_1{*}(\bar\zeta_1{*}\zeta_1)_{+}^{-1}{*}\bar\zeta_k, \quad k=2{\sim}m
$$
For simplicity, denote these by 
$$
\mu, \,\,\tau,\,\, {\eta}_k,\,\,{\bar\eta}_k, \quad k=2{\sim}m,
$$
and let ${\Cal A}$ be the algebra generated by these in  
${\mathcal E}_{2+}({\Bbb C}^{2m})$.

By using 
$$
[\zeta_1, \zeta_1{*}(\bar\zeta_1{*}\zeta_1)_{+}^{-1}]
{=}\zeta_1{*}[\zeta_1,(\bar\zeta_1{*}\zeta_1)_{+}^{-1}]{=}
{-}\zeta_1{*}(\bar\zeta_1{*}\zeta_1)_{+}^{-1}{*}
[\zeta_1,\bar\zeta_1{*}\zeta_1]{*}(\bar\zeta_1{*}\zeta_1)_{+}^{-1}
{=}2\nu{*}(\zeta_1{*}(\bar\zeta_1{*}\zeta_1)_{+}^{-1})^2,
$$ 
we easily see that 
$$
[\mu,\tau]{=}2\mu^2, \quad [\tau,{\eta}_k]{=}\mu{*}{\eta}_k,\quad  
[\tau,\bar{\eta}_k]{=}\mu{*}\bar{\eta}_k,\quad [\eta_k, \bar\eta_{\ell}]{=}\mu\delta_{k\ell},
\quad {\text{others are}}\,\,0.
$$
Hence, this system generates a contact Weyl algebra with a canonical conjugate of
$\mu^{-1}$. Moreover, one may join $\zeta_1$ to ${\Cal A}$ and 
${\Cal A}[\zeta_1]$ is a $\mu$-regulated algebra. 
However, just as in Proposition\,\ref{nongo}, there is no 
standard vacuum $\varpi$ such that $\zeta_1{*}\varpi{=}0$ or
$\varpi{*}\zeta_1{=}0$ holds. This is because $\varpi{*}\zeta_1{=}0$
gives $\varpi{*}{\Cal A}{=}\{0\}$ and  
$\zeta_1{*}\varpi{=}0$ gives $\mu{*}\varpi{=}\infty$.

\section{The case  embedded  Heisenberg  algebra}\label{embedded} 
We saw that the semi-boundedness of $\mu^{-1}$ suffers the existence
of its canonical conjugate, and the localizations by joining 
a divisor or an $*$-inverse of some element defined by integrals 
suffers the existence of standard abstract vacuum.

\medskip
Recall now the embedding, given in \S\,\ref{embedding}
Theorem\,\ref{thm} into  ${\mathcal E}_{1+}({\Bbb C}^{2m{+}2})$.  
This may be viewed as a local normal form of the system mentioned in 
the previous subsection. 
In this case, the element $\rho^2{=}e_*^{2y_0}$ is brought 
from the outside of original Heisenberg algebra ${\Cal H}_{2m}[\mu]$. 
Namely, $\mu^{-1}{=}\h^{-1}e_*^{2y_0}$ is an {\it external energy variable}. 
As $\mu{=}\h e_*^{-2y_0}$, and 
$\tau{=}\frac{1}{2}(e_*^{-2y_0}{*}x_0{+}x_0{*}e_*^{-2y_0})$ is a
canonical conjugate of $\mu^{-1}$, $\tau$ may be viewed as the external time. 

In the extended Weyl algebra
generated by $x_0, y_0$, we denote the vacuum 
$$
\varpi_{00}{=}\lim_{t\to{-}\infty}e^{\frac{1}{2}t}e_*^{t\frac{1}{i\h}x_0{\ctt}y_0}.
$$ 
Note first that $\varpi_{00}$ in not in  
${\mathcal E}_{1+}({\Bbb C}^{2m{+}2})$, but in 
${\mathcal E}_{2+}({\Bbb C}^{2m{+}2})$. 

\bigskip 
The abstract vacuum 
$\widetilde{\varpi}_{s}(\Cal H){=}e_*^{-is\tau}{*}{\widetilde{\varpi}(L_0)}$ 
given in \eqref{Heisenvac} will be called the 
{\bf Heisenberg vacuum}\\
after showing the remarkable property that 
$e_*^{-is\tau}{*}{\widetilde{\varpi}(L_0)}=0$ for $\frac{1}{1{+}2\h s}\leq 0$ 
in the later section (cf. Theorem\,\ref{surprise00}). 
Using $[\mu^{-1}, e_*^{-is\tau}]{=}s\h e_*^{-is\tau}$ 
and $\mu^{-1}{*}\varpi_{00}{=}\frac{1}{\h}\varpi_{00}$,
we have 
$$
\mu{*}e_*^{-is\tau}{*}{\widetilde{\varpi}(L_0)}{=}
\left\{
\begin{matrix}
\frac{\h}{1{+}2\h s}e_*^{-is\tau}{*}{\widetilde{\varpi}(L_0)}, &
\frac{1}{1{+}2\h s}>0\\
 0,& \frac{1}{1{+}2\h s}\leq 0
\end{matrix}
\right.
$$ 
We take the representation space as 
$$
C^{\infty}({\mathbb R}_{s{>}s_0}{\times}{\mathbb
  R}^m){*}\widetilde{\varpi}_{s}(\Cal H),\quad s_0=-\frac{1}{2\h}.
$$
Then, we see $\tilde u_k$ is represented as a multiplication 
$\tilde{u}_k{*}f(s,\tilde{\pmb u}){*}\widetilde{\varpi}_{s}(\Cal H)
{=}
(\tilde{u}_k{*}f(s,\tilde{\pmb u})){*}\widetilde{\varpi}_{s}(\Cal H)$, and 
$$
\mu{*}f(s,\tilde{\pmb u}){*}\widetilde{\varpi}_{s}(\Cal H){=}
\frac{\h}{1{+}2\h s}f(s,\tilde{\pmb u}){*}\widetilde{\varpi}_{s}(\Cal H),
\quad 
\tilde{v}_k{*}f(s,\tilde{\pmb u}){*}\widetilde{\varpi}_{s}(\Cal H){=}
-\frac{\h i}{1{+}2\h s}\partial_kf(s,\tilde{\pmb u})
{*}\widetilde{\varpi}_{s}(\Cal H).
$$

\medskip 
$\mu^{-1}$ and $\tau$ are
canonical conjugate pair. 
$\tau$ is represented by a similar form to  
$X=\frac{1}{i\mu}\sum_{k=1}^{m}{\tilde u}_k{*}{\tilde v}_k$ 
\begin{equation}\label{relativity}
\begin{aligned}
X{*}f(s,{\tilde{\pmb u}}){*}\widetilde{\varpi}_{s}(\Cal H)&=
\frac{\h i}{1{+}2s\h}\sum_{k=1}^m{\tilde u}_k\partial_{\tilde{u}_k}
f(s,{\tilde{\pmb u}}){*}\widetilde{\varpi}_{s}(\Cal H)\\
i\tau{*}f(s,{\tilde{\pmb u}}){*}\widetilde{\varpi}_{s}(\Cal H)&=
\frac{\h}{1{+}2s\h}\sum_{k=1}^m{\tilde u}_k\partial_{\tilde{u}_k}
f(s,{\tilde{\pmb u}}){*}\widetilde{\varpi}_{s}(\Cal H){+}
f(s,{\tilde{\pmb u}}){*}i\partial_s\widetilde{\varpi}_{s}(\Cal H)
\end{aligned}
\end{equation}

In the representation space $\mu$ behaves as $\frac{i}{1{+}2\h s}$.
Tending $s\to\infty$ gives $\mu \to 0$. This gives a classical limit. 
The most remarkable fact is that 
${\rm{Ad}}(e_*^{it\tau})$ {\bf is well-defined only for $t\geq 0$}
(cf. Proposition\,\ref{surprise10}).   
Hence, 
$$
e_*^{it\tau}{*}f(s){*}\widetilde{\varpi}_{s}(\Cal H){=}f(s){*}\widetilde{\varpi}_{s{+}t}(\Cal H)
$$
for only $t{>}0$. In general, 
$$
e_*^{it\mu^{-1}\tau}{*}f(s,\pmb{u}){*}\widetilde{\varpi}_{s}(\Cal H){=}
f(s, E_{a(s)t}(\pmb{u}))\widetilde{\varpi}_{s{+}t}(\Cal H),\quad t\geq
0,\quad a(s){=}\frac{\h}{1{+}2s\h}
$$

\bigskip
\noindent
{\bf Vacuum representation with respect to ${\widetilde{\varpi}_{y_0}(\overline{L}_0)}$}\\
On the other hand, as 
$(x_0{+}i\h){*}e_*^{y_0}{*}\overline{\varpi}_{00}{=}e_*^{y_0}{*}x_0{*}\overline{\varpi}_{00}$, 
the abstract vacuum \eqref{contvac} satisfies 
$\tau{*}{\widetilde{\varpi}_{y_0}(\overline{L}_0)}{=}0$ in generic
ordered expression, and hence 
$$
e_*^{it\tau}{*}\widetilde{\varpi}_{y_0}(\overline{L}_0){=}\widetilde{\varpi}_{y_0}(\overline{L}_0),
\quad t{\in}{\mathbb C}.
$$
As $[i\tau, \mu^{-1}]{=}[i\tau, \mu^{-1}]{=}2$, we have  
$$
[i\tau, f(\mu)]{=}-2\mu^2f'(\mu),\quad [i\tau,{\sqrt\mu}]{=}\frac{1}{\sqrt\mu}
$$

It follows 
$$
i\tau{*}f(\mu^{-1}){*}\widetilde{\varpi}_{y_0}(\overline{L}_0){=}
[i\hat\tau, f(\mu^{-1})]{*}\widetilde{\varpi}_{y_0}(\overline{L}_0){=}
2f'(\mu^{-1}){*}\widetilde{\varpi}_{y_0}(\overline{L}_0).
$$
On the other hand, as
$\tilde{u}_k{=}\frac{\sqrt{\h}}{\sqrt{\mu}}{*}{x}_k$, we have 
$[i\hat\tau, \tilde{u}_k]{=}\mu{*}\tilde{u}_k$, hence 
$$
i\hat\tau{*}f(\tilde{\pmb u}){*}\widetilde{\varpi}_{y_0}(\overline{L}_0){=}
\mu{*}\sum_k\tilde{u}_k\partial_{\tilde{u}_k}f(\tilde{\pmb u}){*}\widetilde{\varpi}_{y_0}(\overline{L}_0).
$$
Let $H{\!o}l({\mathbb C}^{m})$ be the space of 
holomorphic functions of 
$\tilde{u}_1\cdots,\tilde{u}_m$. We set the representation space by  
$$
H{\!o}l({\mathbb C}^{m{+}1})[\mu, \mu^{-1}]]
{*}\widetilde{\varpi}_{y_0}(\overline{L}_0).
$$
Then, we see $\tilde{u}_k$ and $\mu^{-1}$ are represented as multiplications and 

\begin{equation}\label{contactnice}
\tau{*}f(\mu^{-1}, \tilde{\pmb u}){*}\widetilde{\varpi}_{y_0}(\overline{L}_0){=}
i\Big(2\partial_{\mu^{-1}}{+}
\mu{*}\sum_k \tilde{u}_k\partial_{\tilde u_k}\Big)
f(\mu^{-1},\tilde{\pmb u}){*}\widetilde{\varpi}_{y_0}(\overline{L}_0).
\end{equation}
In particular, considering the equation with initial data 
$\mu^{-1}{*}\widetilde{\varpi}_{y_0}(\overline{L}_0)$, 
$$
\frac{d}{dt}e_*^{it\tau}{*}\mu^{-1}{*}\widetilde{\varpi}_{y_0}(\overline{L}_0){=}
e_*^{it\tau}{*}[\hat\tau,\mu^{-1}]{*}\widetilde{\varpi}_{y_0}(\overline{L}_0){=}
2{*}\widetilde{\varpi}_{y_0}(\overline{L}_0),
$$ 
we obtain 
$$
e_*^{it\tau}{*}\mu^{-1}{*}\widetilde{\varpi}_{y_0}(\overline{L}_0){=}
(\mu^{-1}{+}2t){*}\widetilde{\varpi}_{y_0}(\overline{L}_0),\quad
t{\in}{\mathbb C}.
$$
In general 
$$
e_*^{it\mu^{-1}\tau}{*}f(\mu^{-1},\tilde{\pmb u}){*}\widetilde{\varpi}_{y_0}(\overline{L}_0){=}
f(\mu^{-1}(1{+}2t), E_t(\tilde{\pmb u})){*}\widetilde{\varpi}_{y_0}(\overline{L}_0)
$$

\bigskip
This gives also 
$$
e_*^{it\tau}{*}\mu{*}\widetilde{\varpi}_{y_0}(\overline{L}_0){=}
\frac{1}{\mu^{-1}{+}2t}{*}\widetilde{\varpi}_{y_0}(\overline{L}_0)
{=}\frac{\mu}{1{+}2t\mu}{*}\widetilde{\varpi}_{y_0}(\overline{L}_0),\quad
t{\in}{\mathbb C}.
$$
One may take $C^{\infty}({\mathbb R}^m)[\mu^{-1},\mu]]{*}\widetilde{\varpi}_{y_0}(\overline{L}_0)$
as the representation space, where $\mu$ and $\tilde{u}_k$ are
represented by multiplication operator, and 
$$
e_*^{it\mu^{-1}{*}\tau}{*}f(\mu,\tilde{\pmb{u}}){*}\widetilde{\varpi}_{y_0}(\overline{L}_0){=}
f\big(\frac{1}{1{+}2t\mu}, E_t(\tilde{\pmb{u}})\big){*}\widetilde{\varpi}_{y_0}(\overline{L}_0). 
$$
Here, to avoid singularities, we have to treat $\mu$ in a formal
$\mu$-regulated algebra by setting   
$$
\frac{1}{1{+}2t\mu}{=}(1{-}2t\mu{+}(2t\mu)^2{+}(2t\mu)^3{-}\cdots.
$$ 

\subsection{The star-exponential functions  $e_*^{it\tau}{*}F$,  $F{*}e_*^{{-}it\tau}$} 
To give the proof of Theorem\,\ref{surprise00}, we start with a little
general setting. Now, recalling 
$$
{:}\tau{:}_0{=}e^{-2y_0}x_0, \quad x_0{*}e^{-2y_0}{=}e^{-2y_0}{*}(x_0{+}2i\h) 
$$
we consider the $*$-exponential function 
${:}e_*^{ite^{-2y_0}{*}(x_0{+}i\h)}{:}_{_K}$, or $*$-the product 
${:}e_*^{ite^{-2y_0}{*}(x_0{+}i\h)}{*}F{:}_{_K}$.  

The equation for these are  
\begin{equation}\label{eq:12}
\begin{aligned}
\frac{d}{dt}F_t={:}(ie_{*}^{-2y_0}{*}(x_0{+}i\h)){:}_{_K}{*_{_K}}F_t,\quad  F_0={:}F{:}_{_K} \\
\frac{d}{dt}\hat{F}_t
=\hat{F}_t{*_{_K}}{:}(ie_{*}^{-2y_0}{*}(x_0{+}i\h)){:}_{_K},\quad
{\hat F}_0={:}F{:}_{_K}.
\end{aligned}
\end{equation}
As the Hermitian structure is defined, applying the anti-automorphism
$a\to \bar{a}$ to the first one to get 
$$
\frac{d}{dt}\overline{F}_t
=-\overline{F}_t{*_{_K}}{:}(ie_{*}^{-2y_0}{*}(x_0{+}i\h)){:}_{_K},\quad
{\overline F}_0={:}\overline{F}{:}_{_K}.
$$ 

Minding the uniqueness of the real analytic solution, we may assume that $F_t=F_t(x_0,y_0)$
and all computations carry out within these two variables. 

In what follows, all computations execute under the Weyl (K=0) ordered expression. 
If the expression parameter $K$ is restricted to  $K{=}0$ (Weyl ordered expression),
we denote ${:}\,\,{:}_{0}$ in such computation.   
Since 
$$
(x_0{+}i\h){*_{0}}F_t(x_0,y_0){=}
(x_0{+}i\h)F_t(x_0,y_0){-}\frac{i\h}{2}\partial_{y_0}F_t(x_0,y_0),\quad 
(x_0{+}i\h){*_{_{\hat K}}}F_t(x_0,y_0){=}(x_0{+}i\h)F_t(x_0,y_0),
$$
and 
\begin{equation*}
e^{-2y_0}*_0f(x_0, y_0)=e^{-2y_0}f(x_0{-}{i\h}, y_0),\quad
e^{-2y_0}*_{_{\hat K}}f(x_0, y_0)=e^{-2y_0}f(x_0{-}2{i\h}, y_0), 
\end{equation*}
we see that  \eqref{eq:12} turns out to be
\begin{equation}\label{diffequniq}
\frac{d}{dt}F_t(x_0,y_0)= 
 ie^{-2y_0}x_0F_t(x_0{-}{i\h},y_0) 
  +\frac{\h}{2}e^{-2y_0}\rd_{y_0}F_t(x_0{-}i\h, y_0),
\end{equation}
with initial condition $F_0$.

\medskip
Putting $F_t(x_0,y_0)= G_t(x_0,y_0)e^{\frac{2}{i\h}(x_0{+}i\h)y_0}$, the equation
\eqref{diffequniq} becomes 
\begin{equation}
  \label{eq:eqeq2}
\frac{d}{dt}G_t(x_0,y_0)=\frac{\h}{2}e^{-4y_0}\rd_{y_0}G_t(x_0{-}i\h,y_0),\quad 
G_0=F_0(x_0,y_0)e^{-\frac{2}{i\h}(x_0{+}i\h)y_0}.
\end{equation}

\eqref{eq:eqeq2} is not a differential equation, but a differential-difference equation. 

\bigskip
\noindent
{\bf The case $G_t(x_0,y_0)$ is periodic.} 

If $G_t$ is restricted to periodic functions 
$G_t(x_0{-}{i\h},y_0)=G_t(x_0,y_0)$, then \eqref{eq:eqeq2} turns out to be  
\begin{equation*}
\big(\rd_t-\frac{\h}{2}e^{-4y_0}\rd_{y_0}\big)G_t(x_0,y_0)=0,
\end{equation*}
and the solution with the initial data is given by 
\begin{equation*}
G_t(x_0,y_0)= \varphi\bigl(x_0, 2{\h}t{+}e^{4y_0}\bigr),\quad 
G_0(x_0,y_0)= \varphi\bigl(x_0, e^{4y_0}\bigr)
\end{equation*}
by using every holomorphic function $\varphi(u,v)$ such that 
$\varphi(u{+}i\h,v)=\varphi(u,v)$. 
Thus the initial function $G_0(x_0,y_0)$ must satisfy 
the periodical conditions $G_0(x_0{-}i\h,y_0){=}G_0(x_0,y_0)$. 
This is equivalent with $F(x_0{-}i\h,y_0){=}F(x_0,y_0)e^{-2y_0}$. 

Hence ${:}e^{it\tau}{:}_0$ is not defined by itself but it is defined only in the
form ${:}e^{it\tau}{*}F_*{:}_{0}$. That is in the form of solutions of 
equations written under the Weyl ordered expression with the initial
data $F_*$. 

\medskip
For any $\varphi(x_0, y_0)$ satisfying 
$\varphi(x_0{-}i\h, y_0){=}\varphi(x_0, y_0)$ (e.g. $\varphi(y_0)$), 
we set 
$F{=}\varphi(x_0, e^{4y_0})e^{\frac{2}{i\h}(x_0{+}i\h)y_0}$. Then, we see 
\begin{equation}\label{holosol}
{:}e_*^{it\tau}{:}_0{*_0}\varphi(x_0, e^{4y_0} )e^{\frac{2}{i\h}(x_0{+}i\h)y_0}{=}
\varphi(x_0,2\h t{+}e^{4y_0})e^{\frac{2}{i\h}(x_0{+}i\h)y_0}.
\end{equation}
Thus, if $t{\in}{\mathbb C}$, then $\varphi(x_0, y_0)$
must be entire w.r.t. $y_0$.  
Recalling 
$$
{:}\overline{\varpi}_{00}{:}_0{=}2e^{\frac{2}{i\h}x_0y_0},\quad  
{:}f(y_0){*}\overline{\varpi}_{00}{:}_0{=}f(2y_0)2e^{\frac{2}{i\h}x_0y_0},
$$
we see 
$e^{\frac{2}{i\h}(x_0{+}i\h)y_0}{=}{:}e^{y_0}{*}\overline{\varpi}_{00}{:}_0$.

In particular, restricting $\varphi$ to a entire function
$\varphi(z)$ and noting that
$\varphi(e^{2y_0}){=}{:}\varphi(e_*^{2y_0}){:}_0$, 
  \eqref{holosol} is written as
$$
{:}e_*^{it\tau}{*}\varphi(e_*^{2y_0}){*}e_*^{y_0}{*}\overline{\varpi}_{00}{:}_0{=}
{:}\varphi(e_*^{2y_0}{+}2\h t){*}e_*^{y_0}{*}\overline{\varpi}_{00}{:}_0.
$$
Although the computation was done under the Weyl ordered expression,
one may write this without suffix ${:}\,\,{:}_0$, that is, 
\begin{equation}\label{barvacrep}
e_*^{it\tau}{*}\varphi(e_*^{2y_0}){*}e_*^{y_0}{*}
\overline{\varpi}_{00}=\varphi(e_*^{2y_0}{+}2\h t){*}e_*^{y_0}{*}\overline{\varpi}_{00},
\end{equation}
as all terms in r.h.s. can be intertwined to generic expressions.

Recall that $e_*^{y_0}{*}\overline{\varpi}_{00}$ is an abstract vacuum such that 
$$
\overline{\varpi}_{00}{*}y_0{=}0,\quad 
(x_0{+}i\h){*}e_*^{y_0}{*}\overline{\varpi}_{00}=0,\quad
i\tau{*}e_*^{y_0}{*}\overline{\varpi}_{00}=0.
$$ 
Hence, \eqref{barvacrep} may be written as 
$$
e_*^{it\tau}{*}\varphi(e_*^{2y_0}){*}e_*^{-it\tau}{*}e_*^{y_0}{*}
\overline{\varpi}_{00}=
{\rm{Ad}}(e_*^{it\tau})\varphi(e_*^{2y_0}){*}e_*^{y_0}{*}\overline{\varpi}_{00}
{=}\varphi(e_*^{2y_0}{+}2\h t){*}e_*^{y_0}{*}\overline{\varpi}_{00},\quad 
t{\in}{\mathbb C}.
$$
On the other hand, it is easy to see that 
$[i\tau, e_*^{2y_0}]{=}2\h$, $[i\tau, [i\tau, e_*^{2y_0}]]{=}0, \cdots$. This gives 
$$
\exp t{\rm{ad}}(i\tau)\varphi(e_*^{2y_0}){=}\varphi(e_*^{2y_0}{+}2\h t), 
\quad  t{\in}{\mathbb C}
$$
for every entire function $\varphi$. 

\medskip
Applying  these to $e_*^{-2y_0}$, we have 
$$
e_*^{it\tau}{*}e_*^{-2y_0}{*}e_*^{y_0}{*}
\overline{\varpi}_{00}=(e_*^{2y_0}{+}2\h t)^{-1}{*}e_*^{y_0}{*}\overline{\varpi}_{00}
{=}
e_*^{-2y_0}\frac{1}{1{+}2\h t e_*^{-2y_0}}{*}e_*^{y_0}{*}\overline{\varpi}_{00}.
$$
$$
\exp t{\rm{ad}}(i\tau)(e_*^{-2y_0}){=}e_*^{-2y_0}\frac{1}{1{+}2\h t e_*^{-2y_0}}, 
\quad  t{\in}{\mathbb C}.
$$
This is the same to the previous case 
${\widetilde{\varpi}_{y_0}(\overline{L}_0)}$.

\bigskip
\noindent 
{\bf Fourier transform expressions} 

Next we want to remove the periodical condition $G_0(x_0{-}i\h,y_0)=G_0(x_0,y_0)$.
Instead, we assume that $t$ is a real variable and $G_t(x_0,y_0)$ is written as  
\begin{equation*}
G_t(x_0,y_0)=\int_{-\infty}^{\infty} a(t,\xi,y_0)e^{i\xi\,x_0}\dbar\xi, 
\quad G_0(x_0,y_0)=\int_{-\infty}^{\infty} a(0,\xi,y_0)e^{i\xi\,x_0}\dbar\xi 
\end{equation*}
by using tempered distribution $a(t,\xi,y_0)$ of compact support with respect to $\xi$. 

If $x_0$, $y_0$ in  \eqref{eq:diff} are treated as real
variables, then the compactness of the support is removed, but  
$G_t(x_0{-}i\h,y_0)$ must be viewed as the shift of $\xi$
$$
G_t(x_0{-}i\h,y_0)=\int_{-\infty}^{\infty}
a(t,\xi,y_0)e^{i\xi(x_0{-}i\h)}\dbar\xi =
\int_{-\infty}^{\infty}
a(t,\xi,y_0)e^{\h\xi}e^{i\xi x_0}\dbar\xi.
$$
Precisely, $a(t,\xi,y_0)$ is
regarded for every $(t,y_0)$ as an 
$\dS'$-valued function $a(t,y_0)(\xi)$ for every test function
$\psi(\xi)\in \dS$. 
It is remarkable in the Weyl ordered expression that 
the equation \eqref{eq:eqeq2} is changed into the differential equation
\begin{equation}
  \label{eq:diff}
\bigl(e^{-\h\xi}\partial_t {-}\frac{\h}{2}e^{-4y_0}\partial_{y_0}\bigr)a(t,\xi,y_0)= 0
\end{equation}
involving $\xi$ as a parameter, i.e. 
\begin{equation*}
\int\Big(e^{-\h\xi}\rd_t a(t,y_0)(\xi)
  -\frac{\h}{2}e^{-4y_0}\rd_{y_0} a(t,y_0)(\xi)\Big)\psi(\xi)\dbar\xi = 0.
\end{equation*}

\medskip
\eqref{eq:diff} shows that $a(t,\xi,y_0)$ is constant along the real 
vector field $e^{-\h\xi}\partial_t -\frac{\h}{2}e^{-4y_0}\partial_{y_0}$ 
on the $(y_0, t)$-surface.  
That is constant along the integral curves of this vector field. Hence,
the solution is  
\begin{equation}\label{solution-1} 
a(t,\xi,y_0)=\phi(\xi, 2\h t{+}e^{-\h\xi}e^{4y_0}),\quad
a(0,\xi,y_0)=\phi(\xi, e^{-\h\xi}e^{4y_0}) 
\end{equation}
by using a function $\phi(\xi,\eta)$ on ${\mathbb R}^2$ such that 
$f(\xi)=\phi(\xi, 2\h t{+}e^{-\h\xi}e^{4y_0})$ is 
a tempered distribution of compact support.
If we allow $x_0$ to be a real variable, then 
the condition of compact support is relaxed simply as follows: 

\medskip
\noindent
{\bf Condition 1}.  $h(\eta)=\phi(\xi,\eta)$ is a $\mathcal S'$-valued $C^\infty$ 
function such that for every $\eta$, and  for every $(t,y_0)$,   
\begin{equation*}
\phi(\xi, 2{\h t}{+}e^{-\h\xi}e^{4y_0})
\end{equation*}
is also a tempered distribution w.r.t. $\xi$.

\begin{lem}\label{condition}
If $\phi(\xi,\eta)$ is a $C^{\infty}$ function such that 
$|\partial_{\eta}^n\phi(\xi, \eta)\eta^n|<\infty$ for every $n$, then  
Condition 1 is fulfilled. 
\end{lem}

\noindent
{\bf Proof}\,\, As 
$$
\frac{d}{d\xi}\phi(\xi, 2{\h t}{+}e^{-\h\xi}e^{4y_0}){=}
(\partial_{\xi}\phi(\xi, 2{\h t}{+}e^{-\h\xi}e^{4y_0}){+}
\h e^{-\h\xi}e^{4y_0}\partial_{\eta}\phi(\xi, 2{\h t}{+}e^{-\h\xi}e^{4y_0})).
$$ 
Repeating this it is enough to show the boundedness of
$$
(\h e^{-\h\xi}e^{4y_0})^n\partial^n_{\eta}\phi(\xi, 2{\h t}{+}e^{-\h\xi}e^{4y_0}))
$$
This is easily checked. ${}$\hfill $\Box$

\bigskip

So far, the variable $y_0$ has been regarded as a complex variable,
but Condition 1 allows $x_0$ to be a real variable. 
Such an unbalanced situation yields the following:
\begin{thm}\label{singlecondi}
If the initial condition $a(0,\xi,y_0)$ has not periodic property 
$a(0,\xi,y_0){=}a(0,\xi,y_0{\pm}\frac{1}{2}\pi i)$, then
there is no solution of  \eqref{eq:12}. 
This periodical condition is equivalent with 
$G(x_0, y_0{\pm}\frac{1}{2}\pi i){=}G(x_0,y_0)$, and with  
$F(x_0,y_0{\pm}\frac{1}{2}\pi i){=}F(x_0,y_0)e^{\frac{\pi}{\h}x_0}$. 

In particular, the complex $*$-exponential function $e_*^{z\tau}$
cannot exist by itself. It exists only in the 
form $e_*^{z\tau}{*}F$ with initial function $F$ with a
quasi-periodical property $F(x_0,y_0{+}\pi i){=}F(x_0,y_0)e^{\frac{\pi}{\h}x_0}$.
\end{thm}

\medskip
As $t$ and $\xi$ are real variables, it is natural to allow $y_0$ to
be a real variable. Then, \eqref{eq:diff} can be solved without any 
restriction.

\bigskip
To simplify the argument, we put several assumptions in what follows:

\medskip
\noindent
{{\bf Assumption.}\,\, $\h$ is a positive constant. $t$, $\xi$ and $y_0$ 
are real variables.  Under such restrictions,  ${:}e_*^{it\tau}{:}_0$
may be expressed as a function $F_t(x_0,y_0)$ of real variables.

\bigskip

In real variable expressions, it occurs a strange phenomenon. 
Fix $\xi$ arbitrarily. 
If we set $\eta{=}2\h t{+}e^{-\h\xi}e^{4y_0}$, then 
$a(t,\xi,y_0)$ is constant on the curve 
$\eta{=}2\h t{+}e^{-\h\xi}e^{4y_0}$. Thus, if 
$\eta\leq 0$, then $\eta{-}e^{-\h\xi}e^{4y_0}{=}2\h t$ is negative.  
Such a curve does not cross 
the initial line $t=0$. 
As the initial data is given at $t=0$,  
the initial data must be given only when $\eta{>}0$  
by the form
\begin{equation}\label{initial}
a(0,\xi,y_0)=\phi(\xi, \eta), \quad \eta=e^{-\h\xi}e^{4y_0}\quad 
i.e.\,\,\, 4y_0=\h\xi{+}\log\eta.
\end{equation}
\noindent
\unitlength 0.1in
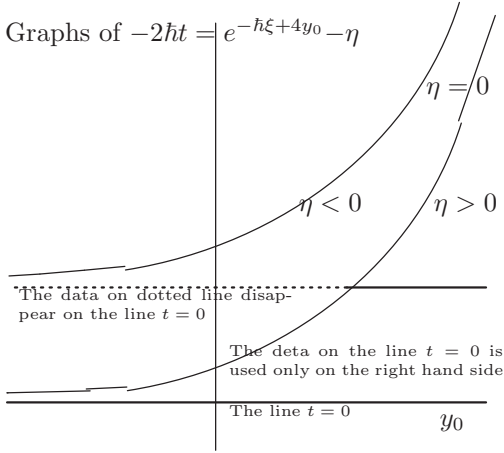
\begin{picture}( 25.3700, 23.3800)(2.1000,-25.7800)
%
\special{pn 8}%
\special{ar 450 258 2200 1940  0.2670092 1.3911711}%
%
\special{pn 8}%
\special{pa 2558 758}%
\special{pa 2748 208}%
\special{fp}%
%
\special{pn 8}%
\special{pa 840 2148}%
\special{pa 630 2158}%
\special{fp}%
\special{pa 650 2168}%
\special{pa 220 2178}%
\special{fp}%
\special{pa 220 2178}%
\special{pa 220 2178}%
\special{fp}%
%
\special{pn 13}%
\special{pa 220 2228}%
\special{pa 2700 2228}%
\special{fp}%
\special{pa 2700 1628}%
\special{pa 1980 1628}%
\special{fp}%
%
\special{pn 13}%
\special{pa 1980 1628}%
\special{pa 230 1628}%
\special{dt 0.045}%
%
\special{pn 8}%
\special{pa 1300 248}%
\special{pa 1300 2478}%
\special{fp}%
\put(24.6000,-23.7800){\makebox(0,0)[lb]{\footnotesize$y_0$}}%
\put(23.9000,-6.4800){\makebox(0,0)[lb]{\footnotesize{$\eta=0$}}}%
\put(17.3000,-12.5800){\makebox(0,0)[lb]{\footnotesize{$\eta<0$}}}%
\put(24.3000,-12.5800){\makebox(0,0)[lb]{\footnotesize{$\eta>0$}}}%
\put(2.1000,-3.6800){\makebox(0,0)[lb]{\footnotesize{Graphs of $-2\h t=e^{-\h\xi{+}4y_0}{-}\eta$}}}%
%
\special{pn 8}%
\special{ar 440 -372 2200 1940  0.2670092 1.3911711}%
%
\special{pn 8}%
\special{pa 830 1518}%
\special{pa 620 1538}%
\special{fp}%
\special{pa 620 1538}%
\special{pa 390 1558}%
\special{fp}%
\special{pa 390 1558}%
\special{pa 230 1568}%
\special{fp}%
\put(13.7000,-23.00){\makebox(0,0)[lb]{\tiny{The line $t=0$}}}%
\put(13.7000,-21.00){\makebox(0,0)[lb]
{\parbox{.2\linewidth}{\tiny{The deta on the line $t=0$ is used only on the right hand side}}}}%
\put(2.8000,-18.0800){\makebox(0,0)[lb]
{\parbox{.2\linewidth}{\tiny{The data on dotted line
      disappear on the line $t=0$}}}}
\end{picture}%
\hfill
\parbox[b]{.6\linewidth}{
Another word, we have 
\begin{prop}\label{unicity}
If $a(0,\xi,y_0){=}a(0,\xi,\frac{1}{4}(\h\xi{+}\log\eta))=0$ for
$\eta>0$, then $a(t,\xi,y_0)=0$ for $\forall t\geq 0$, that is,  
$G_t(x_0,y_0)=0$ for  $\forall t\geq 0$. 
\end{prop}
Given $\phi(\xi, \eta)$, the data at $t=0$ is $\phi(\xi,e^{-\h\xi+4y_0})$
given by using only the values on $\eta>0$, and 
the data at $t=t_0$ is $\phi(\xi,e^{-\h\xi+4y_0}{+}2\h t_0)$.
Some data at $t=0$ disappears in the time $t_0{>}0$, and some data on
$\eta{<}0$ is used at $t_0{<}0$. 
 
Another word {\bf one may choose arbitrary function on the domain} $\eta<0$.
For \eqref{eq:12} {\bf the uniqueness does not hold in the direction $t_0{<}0$}.
}

\medskip
In spite of this, we note the next
\begin{prop}\label{uniqueness}
If one sets $\phi(\xi,\eta)=0$ for $\eta\leq 0$, namely one 
does not use the data for  $\eta\leq 0$, then the 
solution of \eqref{eq:12} is uniquely determined by the 
data at $t=0$. 
\end{prop}

Now, start with an initial ($t{=}0$) data $a(0,\xi,y_0){=}
a(0,\xi,\frac{1}{4}(\h\xi{+}\log\eta))$, and take the 
unique solution $a(s,\xi,\frac{1}{4}(\h\xi{+}\log\eta))$
at $s$ ($s>0$). This is uniquely determined but the data at 
$0{<}\eta{<}2\h s$ is not used in the data at $t=s$.  Now, want to restart using  
the data at $t{=}s$ as the initial data and go back to the line $t=0$. 
Then the original data at $t{=}0$ is not recovered. 
This procedure violates the exponential law, that is, the associativity of 
the one parameter product formula.

Now, integrating the result in Proposition\,\ref{unicity} by $\xi$,
we have   
\begin{prop}\label{nonunique}
For an arbitrary tempered distribution $\phi(\xi,\eta)$ w.r.t $\xi$ 
satisfying Condition 1 and $\phi(\xi,\eta)=0$ for $\eta>0$, the integral 
$G_t(x_0,y_0){=}\int_{\mathbb R}\phi(\xi, 2\h t{+}e^{-\h\xi{+}4y_0})e^{i\xi x_0}\dbar\xi$
satisfies $G_{t}(x_0,y_0){=}0$ for $t{\geq}0$. 
Therefore, $F_t(x_0,y_0){=}G_t(x_0,y_0)e^{\frac{2}{i\h}(x_0{+}i\h)y_0}$
satisfies the equation \eqref{eq:12}, but $F_{t}(x_0,y_0){=}0$ for
$t{\geq}0$.  

If one wants to use the value at $t=t_0{>}0$ as the initial data, 
then ${-}2\h(t{+}t_0){=}e^{-\h\xi{+}4y_0}{+}\eta$ is used instead of 
$-2\h t{=}e^{-\h\xi{+}4y_0}{+}\eta$, where nonvanishing data remains.
\end{prop}

Since the uniqueness ensured only by setting
$\phi(\xi,\eta)=0$ on the area $\eta<0$, we see 
\begin{prop}\label{surprise10}
The adjoint operator 
${\rm{Ad}}(e_*^{it\tau})F=e_*^{it\tau}{*}F{*}e_*^{-it\tau}$ is 
defined only for $t\geq 0$ by using 
$$
e_*^{it\tau}{*}(\overline{e_*^{it\tau}{*}\overline{F}}), \quad t\geq 0.
$$
\end{prop}

Let $\mathcal I(0)$ be the space of all compactly supported 
 tempered distribution
$\phi(\xi,\eta)$ w.r.t. $\xi$.\\ 
Every $\phi(\xi,\eta)\in \mathcal I(0)$ satisfies Condition 1. We set 
$\hat{g}_a(\xi,y_0){=}\phi(\xi,2\h t{+}e^{-\h\xi{+}4y_0})$ for
$t\in{\mathbb R}$ and 
$$
{:}g_t(x_0,y_0){:}_0=\int_{\mathbb R}\hat{g}_t(\xi,y_0)e^{i\xi x_0}\dbar\xi.
$$
Then, the integral
$$
F_t(x_0,y_0){=}
\int_{\mathbb R}
{\hat g}_t(\xi,\frac{1}{4}(\h\xi{+}\log(e^{-\h\xi+4y_0}{+}2\h t))
e^{i\xi x_0}\dbar\xi \,
e^{\frac{2}{i\h}(x_0{+}i\h)y_0} 
$$
is the solution of \eqref{diffequniq} uniquely determined for the
``future'' direction by the data given at $t=a$. Hence one may write 
$$
F_t(x_0,y_0){=}{:}e_*^{it\tau}{:}_{0}{*_0}F_0(x_0,y_0),\quad t\geq 0.
$$
By the uniqueness gives the following:

\begin{prop}\label{keyprop}
The exponential law 
${:}e_*^{i(s{+}t)\tau}{:}_{0}{*_0}F_0(x_0,y_0){=}
{:}e_*^{is\tau}{*}e_*^{it\tau}{:}_{0}{*_0}F_0(x_0,y_0)$ holds for $s, t{>}0$.
\end{prop}

The next one may sound surprising:
\begin{prop}\label{keyprop22}
If the initial $\phi(\xi, \eta)\in \mathcal I(0)$ is compactly supported
w.r.t. $(\xi,\eta)$, then there is $c>0$ such that 
${:}e_*^{it\tau}{:}_{0}{*_0}F_0(x_0,y_0){=}0$ for $t{\geq}c$. 
$F_0(x_0,y_0)$ may be viewed as {\bf an element of a finite time span  of life}.
\end{prop}

\noindent
{\bf Note}\,\,
As $i\tau$ is skew-hermitian, we have the primitive conservation law:
\begin{equation}\label{consevation}
\frac{d}{dt}\overline{F}_t{*}F_t=0. 
\end{equation}
The above result does not against the conservation law
\eqref{consevation}. This implies simply $\overline{F}_t{*}{F}_t=0$. 
Note also $\overline{\varpi}_{00}{*}{\varpi}_{00}=0$ (cf. \cite{ommy6}).

Recall that \eqref{holosol} shows 
${:}e_*^{it\tau}{:}_{0}{*_0}e^{\frac{2}{i\h}(x_0{+}i\h)y_0}{=}e^{\frac{2}{i\h}(x_0{+}i\h)y_0}$.

\subsubsection{Solutions for the initial value $e^{a\frac{1}{i\h}x_0y_0}$} 

By virtue of the restriction to the real variables, one can obtain the 
solution for many initial values.

Consider next ${:}e_*^{it\tau}{:}_{0}{*_0}e^{a\frac{1}{i\h}x_0y_0}$ 
for general $a\in{\mathbb R}$.
Noting that $e^{(a-2)\frac{1}{i\h}x_0y_0}{=}
\int_{\mathbb R}\delta(\xi{+}\frac{a{-}2}{\h}y_0)e^{i\xi x_0}\dbar\xi$,
we have the initial data 
$$
e^{a\frac{1}{i\h}x_0y_0}{=}\int
e^{-2y_0}\delta(\xi{+}\frac{a{-}2}{\h}y_0)e^{i\xi x_0}\dbar\xi e^{\frac{2}{i\h}(x_0{+}i\h)y_0}.
$$
Hence the initial data $\phi(\xi,\eta)$ in \eqref{initial} is given by 
$$
\phi(\xi,\eta)=e^{-2y_0}\delta(\xi{+}\frac{a{-}2}{\h}y_0){=}
e^{-\frac{1}{2}(\h\xi{+}\log\eta)}
\delta((1{+}\frac{a{-}2}{4})\xi{+}\frac{a{-}2}{4\h}\log\eta),
$$
plugging $y_0=\frac{1}{4}(\h\xi{+}\log(\eta))$. As $\delta(x)$ is
supported only on $x=0$, the support of $\phi(\xi,\eta)$ is given as 
the curve $(\frac{a{+}2}{4})\xi{+}\frac{a{-}2}{4\h}\log\eta=0$ in
the $(\xi,\eta)$ plane:

\noindent
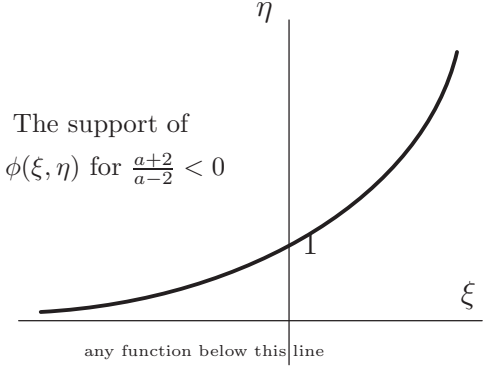
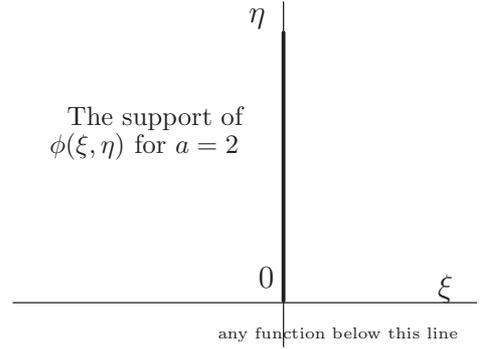
\unitlength 0.1in
\begin{picture}( 64.8000, 19.6000)(  0.6000,-19.3000)
%
\special{pn 20}%
\special{ar 50 40 2380 1620  0.1594132 1.4890119}%
%
\special{pn 8}%
\special{pa 130 1700}%
\special{pa 2530 1700}%
\special{fp}%
\special{pa 1530 1930}%
\special{pa 1530 130}%
\special{fp}%
\put(24.2000,-16.4000){\makebox(0,0)[lb]{$\xi$}}%
\put(13.6000,-1.4000){\makebox(0,0)[lb]{$\eta$}}%
\put(0.6000,-10.0000){\makebox(0,0)[lb]{\footnotesize{$\phi(\xi,\eta)$ for $\frac{a{+}2}{a{-}2}<0$}}}%
\put(16.0000,-13.5000){\makebox(0,0)[lb]{$1$}}%
\put(4.7000,-19.0000){\makebox(0,0)[lb]{\tiny{any function below this line}}}%
\put(1.0000,-7.4000){\makebox(0,0)[lb]{\footnotesize{The support of}}}%
%
\special{pn 8}%
\special{pa 4140 1700}
\special{pa 6540 1700}%
\special{fp}%
\special{pa 5540 1930}%
\special{pa 5540 130}%
\special{fp}%
\put(54.1000,-16.2000){\makebox(0,0)[lb]{$0$}}%
\put(52.,-19.0000){\makebox(0,0)[lb]{\tiny{any function below this line}}}%
\put(53.6000,-2.7000){\makebox(0,0)[lb]{$\eta$}}%
\put(63.4000,-17.0000){\makebox(0,0)[lb]{$\xi$}}%
%
\special{pn 20}%
\special{pa 5540 1690}%
\special{pa 5540 290}%
\special{fp}%
\put(44.2000,-8.0000){\makebox(0,0)[lb]{\footnotesize{The support of}}}%
\put(43.3000,-9.5000){\makebox(0,0)[lb]{\footnotesize{$\phi(\xi,\eta)$ for $a=2$}}}%
\end{picture}%

\medskip

But this gives
$\phi(\xi,\eta)$ only on ``half'' area $\eta>0$. 

\medskip
\fbox{\parbox{.9\linewidth}{To obtain the uniquness, we set
$\phi(\xi,\eta)=0$ for $\eta{<}0$. General solution is obtained by 
adding arbitrary $\phi(\xi,\eta)$ supported on $\eta<0$.}
}

\bigskip
By this convention, the solution is obtained by replacing $\eta$ by
$2\h t{+}e^{-\h\xi{+}4y_0}$ providing \\
$2\h t{+}e^{-\h\xi{+}4y_0}{>}0$ and 
$\eta{=}0$ if $2\h t{+}e^{-\h\xi{+}4y_0}{\leq}0$.

\begin{equation}\label{setnice}
{:}e_*^{it\tau}{:}_0{*}_0e^{a\frac{1}{i\h}x_0y_0}=
\int e^{-\frac{1}{2}(\h\xi{+}log(e^{4y_0{-}\h\xi}{+}2\h t))}
\delta\Big((\frac{a{+}2}{4})\xi{+}\frac{a{-}2}{4\h}
\log(e^{4y_0{-}\h\xi}{+}2\h t)\Big)e^{i\xi x_0}\dbar\xi\,\,e^{\frac{2}{i\h}(x_0{+}i\h)y_0}.
\end{equation}

\bigskip
\noindent
{\bf The case $a{=}-2$}.\,\, This is the case that we want for the
proof of Theorem \,\ref{surprise00}, where 
$\phi(\xi,\eta)=e^{-\frac{1}{2}(\h\xi{+}\log\eta)}\delta(-\frac{1}{\h}\log(\eta))$.

\noindent
\setlength{\unitlength}{2pt}
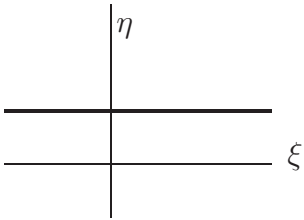
\begin{picture}(50,40)(0,-10)
\thinlines
\put(5,10){\line(1,0){50}}
\put(25,0){\line(0,1){40}}
\thicklines
\put(5,20){\line(1,0){50}}
\put(26,35){$\eta$}
\put(58,10){$\xi$}
\end{picture}
\hfill 
\parbox[b]{.67\linewidth}
{
 That is, the support
of $\phi(\xi,\eta)$ is the line $\eta=1$ in the $(\xi,\eta)$ plane. 
 Setting 
$\xi'={-}\frac{1}{\h}\log(e^{-\h\xi{+}4y_0}{+}2\h t)$, we see that  
if $1{-}2\h t\leq 0$, then  $\xi'$ cannot be $0$.  
Thus, $\delta(\xi')=0$ for $1{-}2\h t\leq 0$. The solution may be written as 
$$
{:}e_*^{it\tau}{:}_0{*}_0e^{-2\frac{1}{i\h}x_0y_0}=
\left\{
\begin{matrix}
0&2\h t\geq 1\\
e^{{-}\frac{\h}{2}(\xi{-}\xi')}\int\delta(\xi')e^{i\xi x_0}\dbar\xi\,\,e^{\frac{2}{i\h}(x_0{+}i\h)y_0}&
2\h t<1\\
\end{matrix}
\right.
$$
}

$\xi'=0$ is at
$\h\xi=4y_0{-}\log(1-2\h t)$, i.e. 
$e^{-\frac{1}{2}\h\xi}|_{\xi'=0}=\sqrt{1{-}2\h t}\,e^{-2y_0}$. It follows 
$$
\frac{d\xi'}{d\xi}\Big|_{\xi'=0}
=\frac{e^{-\h\xi}e^{4y_0}}{e^{-\h\xi}e^{4y_0}{+}2\h t}=1{-}2\h t.
$$

Thus,
\begin{equation}\label{vacuumobt}
{:}e_*^{it\tau}{:}_0{*}_0e^{-2\frac{1}{i\h}x_0y_0}{=}
\left\{
\begin{matrix}
0&2\h t\geq 1\\
\frac{1}{\sqrt{1{-}2\h t}}e^{-i\frac{x_0}{\h}\log(1{-}2\h t)}e^{-\frac{2}{i\h}x_0y_0}&
2\h t{<}1\\
\end{matrix}
\right.
\end{equation}
Recall the vacuum
${:}\varpi_{00}{:}_0{=}2e^{-2\frac{1}{i\h}x_0y_0}$ and the formula 
${:}f(x_0){:}_0{*_0}{:}\varpi_{00}{:}_0=f(2x_0){:}\varpi_{00}{:}_0$. As
the r.h.s. is defined on generic ordered
expression, Proposition\,\ref{judge} gives
that \eqref{vacuumobt} is rewritten without ${:}\,\,{:}_0$ as a
rather surprising form 

\begin{thm}\label{surprise00} 
In generic ordered expression, we have 
$$
e_*^{it\tau}{*}\frac{1}{2}\varpi_{00}{=}
\left\{
\begin{matrix}
  0&2\h t\geq 1\\
\frac{1}{\sqrt{1{-}2\h t}} e_*^{\frac{x_0}{2i\h}\log(1{-}2\h t)}
{*}\frac{1}{2}\varpi_{00}&2\h t{<}1.\\ 
\end{matrix}
\right.
$$
\end{thm}

It is not hard to see that 
$$
\sum_{n{=}0}^{\infty}\frac{1}{n!}(\frac{t}{\h})^n\tau_*^n{*}\varpi_{00}{=}
\sum_{n{=}0}^{\infty}\frac{1}{n!}(\frac{t}{\h})^n\frac{i\h}{x_0}(\frac{x_0}{i\h})_{n+1}{*}\varpi_{00}
$$
without $\mu$.

Hence 
\eqref{vacuumobt} is the ``time evolution'' of the vacuum.  
This is real analytic for $t{<}\frac{1}{2\h}$ and the exponential
law holds on this domain. 
However, Proposition\,\ref{nonunique} shows that solutions are not
unique on 
$\eta=e^{4y_0}{+}2\h t{<}0$ depending on arbitrary functions. 

\subsubsection{The case  that initial date is $\overline{\varpi}_{00}$} 
Consider now the case $a{=}2$. Note first that 
${:}\overline{\varpi}_{00}{:}_0=2e^{\frac{2}{i\h}x_0y_0}$. Hence this
case gives the vacuum representation w.r.t. $\overline{\varpi}_{00}$.
If this is the case, the property of the    
delta function gives 
\begin{equation}\label{confirm}
{:}e_*^{it\tau}{*}\frac{1}{2}\overline{\varpi}_{00}{:}_0=
 e^{-\frac{1}{2}log(e^{4y_0}{+}2\h t)}
e^{\frac{2}{i\h}(x_0{+}i\h)y_0}{=}
\frac{1}{\sqrt{e^{4y_0}{+}2\h t}}e^{\frac{2}{i\h}(x_0{+}i\h)y_0}{=}
\frac{1}{\sqrt{1{+}2\h te^{-4y_0}}}e^{\frac{2}{i\h}x_0y_0}
\end{equation}
This is the case $\varphi(z){=}(\sqrt{z})^{-1}$ in
\eqref{atodetukau}. 
It looks that there is a double branched singularities.
Here we have to consider the treatment of $\sqrt{\,\,\,}$. 
 
By noting that 
${:}(\sqrt[*]{1{+}2\h te_*^{-2y_0}})^{-1}_*{:}_0
{=}\frac{1}{\sqrt{1{+}2\h te^{-2y_0}}}$ 
and  
$f(y_0){*_0}{:}\overline{\varpi}_{00}{:}_0{=}f(2y_0){:}\overline{\varpi}_{00}$,
\eqref{confirm} is rewritten as 
$$
{:}e_*^{it\tau}{*}{:}\overline{\varpi}_{00}{:}_0=
{:}\big(\sqrt[*]{1{+}2\h te_*^{-2y_0}}\big)_*^{-1}{:}_0{*_0}{:}\overline{\varpi}_{00}{:}_0,
$$

\medskip
Compare this with the result \eqref{atodetukau00}. 
As real variables are considered, we have to set $\phi(\xi,\eta)=0$ for     
$\eta=e^{4y_0}{+}2\h t<0$ by the comment between
Proposition\,\,\ref{uniqueness} and Proposition\,\,\ref{nonunique} 
to make the solution unique. 
Thus, the solution might be written by the discontinuous function  
$$
{:}e_*^{it\tau}{*}\frac{1}{2}\overline{\varpi}_{00}{:}_0=
\big(\frac{1}{\sqrt{1{+}2\h te^{-4y_0}}}\big)_{re}e^{\frac{2}{i\h}x_0y_0},\quad 
\big(\frac{1}{\sqrt{1{+}2\h te^{-4y_0}}}\big)_{re}=
\left\{
\begin{matrix}
\frac{1}{\sqrt{1{+}2\h te^{-4y_0}}}  &e^{4y_0}{+}2\h t>0\\
 0& e^{4y_0}{+}2\h t<0.
\end{matrix}
\right.
$$ 

\noindent
{\bf Note}.\,\,What is moving by the equation \eqref{eq:12} is not a
point, but a function. Hence the singular set $2\h t{+}e^{4y_0}{=}0$
is {\it not} a singular point of the equation \eqref{eq:12}. It is
only that the initial smooth function is changed into a function 
involving a singular point depending on $t$.  

\bigskip
However, if we think these under formal power series of $\h$, then 
$\frac{1}{\sqrt{1{+}2\h te^{-4y_0}}}$ makes sense as a formal power
series of $\h e^{-4y_0}$. 
Indeed, using $x_0{*}e_*^{-2ny_0}{=}e_*^{-2ny_0}{*}(x_0{+}2ni\h)$,
we have in generic ordered expression that 
\begin{equation}\label{formalsol}
\sum_{k=0}^{\infty}\frac{(it)^k}{k!}{:}\tau_*^k{*}\overline{\varpi}_{00}{:}_{_K}
{=}{:}\Big(\sqrt[*]{1{+}2\h te_*^{-2y_0}}\Big)_*^{-1}{*}\overline{\varpi}_{00}{:}_{_K}{=}
{:}\frac{1}{\sqrt[*]{1{+}2t\mu}}{*}\overline{\varpi}_{00}{:}_{_K}
\end{equation}
where the r.h.s. stands for the Taylor series of $\mu$.  

\begin{prop}
Although
$e_*^{it\tau}{*}e_*^{y_0}{*}\overline{\varpi}_{00}{=}e_*^{y_0}{*}\overline{\varpi}_{00}$
holds in generic ordered expression, 
$e_*^{it\tau}{*}\overline{\varpi}_{00}$ is defined only in a formal 
$\mu$-regulated algebra.
\end{prop}

\bigskip
On the other hand, taking the hermitian conjugate by using the Hermitian structure, we have
\begin{equation}\label{cmplxcnj}
\varpi_{00}{*}e_*^{y_0}{*}e_*^{-it\tau}=\varpi_{00}{*}e_*^{y_0}.
\end{equation}
The combination of \eqref{cmplxcnj} and \eqref{vacuumobt} 
gives under generic ordered expression that 
\begin{equation}\label{assbroken}
\begin{aligned}
(\varpi_{00}{*}e_*^{y_0}{*}e_*^{-it\tau}){*}\varpi_{00}
&{=}(\varpi_{00}{*}e_*^{y_0}){*}\varpi_{00}
 \\
\varpi_{00}{*}e_*^{y_0}{*}(e_*^{-it\tau}{*}\varpi_{00})
&{=}\varpi_{00}{*}e_*^{y_0}{*}(\frac{1}{\sqrt{1{+}2\h t}}{*}\varpi_{00}),
\end{aligned}
\end{equation}
hence the associativity is broken. This is not a contradiction,
because these are elements of ${\mathcal E}_{2+}({\mathbb C}^{2m})$.

The reason seems that 
$\varpi_{00}{*}e_*^{y_0}$ is a vacuum different from $\varpi_{00}$. 

\bigskip

On the contrary, in the formal level we see the associativity holds: 
The combination of the hermitian conjugate of \eqref{formalsol} and \eqref{vacuumobt} gives by 
formal power series of $\mu$
$$
\begin{aligned}
{:}(\varpi_{00}{*}e_*^{-it\tau}){*}\varpi_{00}{:}_0&=
\Big(2e^{-\frac{2}{i\h}x_0y_0}\frac{1}{\sqrt{1{+}2\h te^{-4y_0}}}\Big){*_0}{:}\varpi_{00}{:}_{0}
{=}\Big({:}\varpi_{00}{*}(\sqrt[*]{1{+}2\h te^{-2y_0}})_*^{-1}{:}_0\Big){*_0}{:}\varpi_{00}{:}_{0} \\
{:}\varpi_{00}{*}(e_*^{-it\tau}{*}\varpi_{00}){:}&=
{:}\varpi_{00}{:}_0{*_0}\Big(\frac{1}{\sqrt{1{+}2\h t}}
e^{-i\frac{x_0}{\h}\log(1{+}2\h t)}2e^{-\frac{2}{i\h}x_0y_0}\Big)\\
&\qquad\qquad
{=}{:}\varpi_{00}{:}_0{*_0}\Big(\frac{1}{\sqrt{1{+}2\h t}}
e^{-i\frac{x_0}{2\h}\log(1{+}2\h t)}{*_0}{:}\varpi_{00}{:}_0\Big)
\end{aligned}
$$
by using 
${:}\varpi_{00}{*}y_0{:}_0{=}{:}\varpi_{00}{:}_02y_0$,
${:}x_0{*}\varpi_{00}{:}_0{=}{:}\varpi_{00}{:}_02x_0$. 
At a first glance, these violate the associativity, but in fact 
as $y_0{*}\varpi_{00}{=}0{=}\varpi_{00}{*}x_0$, we have the associativity 
\begin{equation}\label{justext}
{:}(\varpi_{00}{*}e_*^{-it\tau}){*}\varpi_{00}{:}_0{=}
\frac{1}{\sqrt{1{+}2h t}}{:}\varpi_{00}{:}_{0}
{=}{:}\varpi_{00}{*}(e_*^{-it\tau}{*}\varpi_{00}){:}_0, \quad
2\h t{\geq}-1. 
\end{equation}

\subsubsection{Solution for the initial value 1}\label{1.2.2} 
It is interesting to investigate the solution with the initial value
1, but in this subsection, by technical reason, all computation are restricted under the Weyl
ordered ($K=0$) expression.
 
Note that the initial condition $F_0=1$ corresponds to 
$\phi(\xi,y_0)=e^{-2y_0}\delta(\xi{-}\frac{2}{\h}y_0)$, as
$$
G_0(x_0,y_0)=e^{-\frac{2}{i\h}(x_0{+}i\h)y_0}=
\int_{-\infty}^{\infty}e^{-2y_0}\delta(\xi{-}\frac{2}{\h}y_0)
e^{i\xi\,x_0}\dbar\xi.
$$
Since this is the case $a=0$, \eqref{setnice} gives that the solution obtained by 
putting ${\phi}(\xi,\eta)=0$ for $\eta<0$ is   
\begin{equation*}\label{starexp}
{:}e_*^{it\tau}{:}_0=
\int_{-\infty}^{\infty}\!\!
e^{-\frac{1}{2}(\h\xi{+}\log(e^{-\h\xi{+}4y_0}{+}2\h t))}
\delta\Big(\frac{1}{2}\xi{-}\frac{1}{2\h}\log(e^{-\h\xi{+}4y_0}+2\h t)\Big)
e^{i\xi\,x_0}\dbar\xi\,e^{\frac{2}{i\h}(x_0{+}i\h)y_0}.
\end{equation*}
This is the unique real analytic solution and $e^{-\h\xi{+}4y_0}{+}2\h t>0$
is satisfied.

Note that  $\delta\Big(\frac{1}{2}\xi{-}\frac{1}{2\h}\log(e^{-\h\xi{+}4y_0}+2\h t)\Big)$ 
is supported only on the point 
$\frac{1}{2}\xi{-}\frac{1}{2\h}\log(e^{{-}\h\xi{+}4y_0}{+}2\h t){=}0$. Moreover, 
$$
\xi'=\frac{1}{2}\xi{-}\frac{1}{2\h}\log(e^{{-}\h\xi{+}4y_0}{+}2\h t)
$$  
is easily inverted as 
$$
\xi=\frac{1}{\h}\log\Big(\sqrt{(\h te^{4\h\xi'})^2{+}e^{4y_0}e^{2\h\xi'}}+\h te^{2\h\xi'}\Big).
$$
Hence 
$$
e^{\h\xi}\big|_{\xi'{=}0}=
e^{4y_0-\h\xi}\big|_{\xi'{=}0}{+}2\h t{=}\sqrt{(\h t)^2{+}e^{4y_0}}{+}\h t,\quad 
\frac{d\xi'}{d\xi}=\frac{\sqrt{(\h t)^2{+}e^{4y_0}}}{\sqrt{(\h t)^2{+}e^{4y_0}}{+}\h t}.
$$
Thus, the change of variable of integration gives 
$$
{:}e_*^{it\tau}{:}_0=
\int_{-\infty}^{\infty}\!\!
e^{-\frac{1}{2}(\h\xi{+}\log(e^{4y_0{-}\h\xi}{+}2\h t))}\delta(\xi')
e^{i\xi x_0}
\frac{d\xi}{d\xi'}\dbar\xi'\,e^{\frac{2}{i\h}(x_0{+}i\h)y_0}
=\int_{-\infty}^{\infty}\!\!\Big(e^{-\h\xi}e^{i\xi x_0}
\frac{d\xi}{d\xi'}\Big)\Big|_{\xi'=0}
\dbar\xi'\,e^{\frac{2}{i\h}(x_0{+}i\h)y_0}
$$ 
as $\delta(\xi')$ is supported only on $\xi'=0$. 
Hence we obtain 
\begin{equation*}
{:}e_*^{it\tau}{:}_0={:}e_*^{ite_*^{-2y_0}{*}(x_0{+}ih)}{:}_0{=}
\frac{1}{\sqrt{(\h t)^2{+}e^{4y_0}}}
e^{\frac{ix_0}{\h}\log(\sqrt{(\h t)^2{+}e^{4y_0}}{+}\h t)}
e^{\frac{2}{i\h}(x_0{+}i\h)y_0}
\end{equation*}
As $\sqrt{(\h t)^2{+}e^{4y_0}}{+}\h t>0$ always, this is the unique real
analytic solution with initial condition $1$. We denote this by 
\begin{equation}\label{analyticsol}
\begin{aligned}
{:}\exp_*(it\tau){:}_0&=
\frac{1}{\sqrt{(\h t)^2{+}e^{4y_0}}}
e^{\frac{ix_0}{\h}\log(\sqrt{(\h t)^2{+}e^{4y_0}}{+}\h t)}
e^{\frac{2}{i\h}(x_0{+}i\h)y_0}\\
&=\frac{1}{\sqrt{1{+}(e^{-2y_0}\h t)^2}}
e^{\frac{ix_0}{\h}\log(\sqrt{1{+}(e^{-2y_0}\h t)^2}{+}e^{-2y_0}\h t)}, 
\quad t\in{\mathbb R}\\
\end{aligned}
\end{equation}
This may be rewritten as 
$$
{:}\exp_*(it\tau){:}_0{=}
\frac{1}{\sqrt{1{+}(\mu t)^2}}
e^{\frac{ix_0}{\h}\log(\sqrt{1{+}(\mu t)^2}{+}\mu t)}.
$$
Note also that 
$$
(\sqrt{1{+}(\mu t)^2}{+}\mu t)(\sqrt{1{+}(\mu t)^2}{-}\mu t){=}1,
\quad 
(\sqrt{1{+}(\mu t)^2}{+}\mu t){+}(\sqrt{1{+}(\mu t)^2}{-}\mu t){=}\sqrt{1{+}(\mu t)^2}.
$$

This expression has no singularity and it is easy to check 
$$
\overline{{:}\exp_*(it\tau){:}_0}={:}\exp_*(-it\tau){:}_0.
$$
Note that ${:}i\tau{:}_0=ie^{-2y_0}x_0$ to check 
${:}\frac{d}{dt}\big|_{t=0}e_*^{it\tau}{:}_0={:}i\tau{:}_0$.

\medskip
If an arbitrary $\tilde\phi(\eta)$  supported on the
domain $\eta<0$ is considered, then 
\begin{equation}\label{starexp3}
{:}e_*^{it\tau}{:}_0=
{:}\exp_*(it\tau){:}_0
+\int_{-\infty}^{\infty}\tilde\phi(2{\h t}{+}e^{-\h\xi}e^{4y_0})e^{i\xi x_0}\dbar\xi 
\,e^{\frac{2}{i\h}(x_0{+}i\h)y_0}
\end{equation}
and solutions are not unique in the past direction. 

By these calculations and the uniqueness for future direction, we must have 
the exponential law 
$$
{:}\exp_*(is\tau){:}_0{*_0}{:}\exp_*(it\tau){:}_0={:}\exp_*(i(s{+}t)\tau){:}_0,\quad
s, t \geq 0
$$
holds, and if ${*_0}F$ in the r.h.s. is welldefined then   
$$
{:}\exp_*(it\tau){:}_0{*_0}F=
\frac{1}{\sqrt{(\h t)^2{+}e^{4y_0}}}
e^{\frac{ix_0}{\h}\log(\sqrt{(\h t)^2{+}e^{4y_0}}{+}\h t)}
e^{\frac{2}{i\h}(x_0{+}i\h)y_0}{*_0}F
$$
is the solution of initial data $F$.

\bigskip
\noindent
{\bf Computational conjecture}.  

By the observation above, one must have the identities 
$$
\begin{aligned}
{:}e_*^{-it\tau}{*}\frac{1}{2}\varpi_{00}{:}_0{=}
{:}e_*^{-it\tau}{:}_0{*_0}e^{-\frac{2}{i\h}x_0y_0}
&{=}
\frac{1}{\sqrt{(\h t)^2{+}e^{4y_0}}}
e^{\frac{ix_0}{\h}\log(\sqrt{(\h t)^2{+}e^{4y_0}}{-}\h t)}
e^{\frac{2}{i\h}(x_0{+}i\h)y_0}{*_0}e^{-\frac{2}{i\h}x_0y_0}\\
&{=}
\frac{1}{\sqrt{1{+}2\h t}}e^{-i\frac{x_0}{\h}\log(1{+}2\h t)} 
e^{-\frac{2}{i\h}x_0y_0},\quad -1{<}2\h t.
\end{aligned}
$$
But it is not easy to check these identities by direct calculations.  

It seems possible to prove directly that in some restricted
expressions  
$$
\begin{aligned}
{:}e_*^{it\tau}{*}\frac{1}{2}\overline{\varpi}_{00}{:}_{_K}{=}
&
\frac{1}{\sqrt{(\h t)^2{+}e^{4y_0}}}
e^{\frac{ix_0}{\h}\log(\sqrt{(\h t)^2{+}e^{4y_0}}{+}\h t)}
e^{\frac{2}{i\h}(x_0{+}i\h)y_0}{*_{_K}}e^{\frac{2}{i\h}x_0y_0}\\
&{=}
\frac{1}{\sqrt{1{+}2\h te^{-4y_0}}}e^{\frac{2}{i\h}x_0y_0},\quad e^{4y_0}{+}2\h t>0.
\end{aligned} 
$$

\section{Berezin algebra embedded in the  extended Weyl algebra} 

As it is well known, Berezin operators give operator representations 
of algebras which may be viewed as the quantization 
of the structure of the Poincar{'e} unit disk, 
where the central element corresponding to $\h$ is
represented by an operator. Note that the notion of ${*}$-product is
not used, but as it is naturally expected, the $*$-product that we used in this
note is redefined by using the operator product.
In this section, we give an embedding of Berezin algebra into the
extended Weyl algebra by tracing \cite{B}.

\subsection{Algebra on the unit disk $D$.} \,\,\,
To define Berezin operators we first remark the following:
\begin{lem}\label{Cdelta}  
For every $($anti-$)$ holomorphic function $g$ on $D$ with suitable
growth condition so that the integrals converge, we have for $s{>}-1$    
$$
\frac{(s+1)i}{2\pi}\iint_{D} g(v)(1-v\bar v)^{s}dvd\bar v 
=g(0)=\frac{(s+1)i}{2\pi}\iint_{D} 
g(\bar v)(1-v\bar v)^{s}dvd\bar{v},\quad 
$$
\end{lem}
\par\noindent
{\bf Proof\,\,}
By the Taylor expansion of $g(v)$ at $0$ 
and noting the integral vanishes by 
the integration $\int e^{ik\theta}d\theta$ for $k\not=0$, we have only 
to compute as follows:
$$
\frac{i}{2}\iint_{D}(1-v\bar v)^{s}dvd\bar v =
 2\pi \int_0^1(1-r^2)^{s}rdr = \pi\int_0^1 (1-t)^{s}dt= \frac{\pi}{s+1}.  
$$ 
${}$ \hfill $\Box$ 

\medskip
By the definition of the beta function, we see also the following:
\begin{equation}
  \label{eq:Beta}
\frac{i}{2}\!\iint_{D}(v\bar v)^m(1-v\bar v)^{s} dvd\bar v{=}
2\pi \int_0^1r^{2m}(1-r^2)^{s} rdr 
{=}\pi B(m+1,s+1), \quad {\rm{Re}}\,m{+}1, \,\,{\rm{Re}}\,s{+}1>0 
\end{equation}

Thus, setting 
$$
\Phi_{s}(w,v,\bar v)=
\frac{1}{\pi}\sum_{m=0}^\infty
\frac{1}{B(m+1,s+1)}(1-v\bar v)^s (w\bar v)^m 
=\frac{s+1}{\pi}\frac{(1-v\bar v)^s}{(1-w\bar v)^{s+2}}
$$
we have the identity
\begin{equation}
  \label{eq:identity}
  f(w)= \int_D f(v)\Phi_{s}(w, v, \bar v)dv d\bar v 
= \frac{s+1}{\pi}\int_D f(v)
   \left(\frac{1-v\bar v}{1-w\bar v}\right)^{s+2}dm(v,\bar v)     
\end{equation}
where $dm(v,\bar v)= \frac{i}{2(1-v\bar v)^2}dvd\bar v$ is the volume 
element on the Poincar\'e unit disk. 

Let $\Cal H^{s}(D)$ be the space of all holomorphic functions on $D$
such that 
$$
\int_D |f|^2(1-v\bar v)^{s+2}dm(v,\bar v)<\infty.
$$  

It is known that 
$e_{(s)}^m=(\frac{\Gamma(s+m+2)}{\pi\Gamma(m+1)\Gamma(s+1)})^{1/2}w^m$  
form an orthonormal basis of $\Cal H^{s}(D)$. Note that $s{>}-1$ is
crucial to obtain Hilbert spaces, and no vacuum appears here.
  
Consider the linear space bundle and the space of all smooth sections  
$$
\coprod_{s>-1}\Cal H^{s}(D),\quad \varGamma(\coprod_{s>-1}\Cal H^{s}(D)).
$$

For a function $a(w,\bar w)$ on $D$,  
 the Berezin operator $P_s(a)$ on $\Cal H^s(D)$ is defined by  
\begin{equation}
  \label{eq:identity2}
  (P_s(a)f)(w)=\frac{(s{+}1)i}{2\pi}\int_D a(v,\bar v)f(v)
  \Phi_s(w,v,\bar v)dv d\bar v 
   = \frac{(s+1)}{\pi}
   \int_D a(v,\bar v)f(v)
    \left(\frac{1-v\bar v}{1-w\bar v}\right)^{s+2}dm(v,\bar v).    
\end{equation}
Let $P_s(a)e_{(s)}^m = \sum_k A_{m,k}e_{(s)}^k.$ Then, we have 

\begin{equation}
  \label{eq:mat1}
  A_{m,k}= \frac{1}{\pi}\int_{D}a(v,\bar v)v^m\bar v^k(1-v\bar v)^{s}dvd\bar v
     {\scriptstyle(\frac{\Gamma(s+m+2)}{\Gamma(m+1)\Gamma(s+1)})^{1/2}}
     {\scriptstyle(\frac{\Gamma(s+k+2)}{\Gamma(k+1)\Gamma(s+1)})^{1/2}}
\end{equation}
Hence the operator norm $||P_s(a)||$ is given by 
$$
||P_s(a)||^2=(s+1)^2\iint a(v,\bar v)\overline{a(w,\bar w)}
\left(\frac{1-v\bar v}{1-v\bar w}\frac{1-w\bar w}{1-w\bar v}\right)^{s+2}
      dm(v,\bar v)dm(w,\bar w) 
$$

It is easy to see that
 \begin{equation}
 \label{eq:Berezin}
\begin{aligned}[m]
(P_s(\phi(w))f)(w)=&\int_D \phi(v)f(v)\Phi_s(w,v, \bar v)dv d\bar v 
    = \phi(w)f(w),\\ 
(P_s(\frac{\bar w}{1-w\bar w})f)(w)=& 
  \int_D\frac{\bar v}{1-v\bar v}f(v)\Phi_s(w,v,\bar v)dv d\bar v 
   = \frac{1}{s}f'(w)
\end{aligned}
 \end{equation}
The second one is defined only for $s>0$. 
We define the product $*= *_s$ by the operator product, i.e. 
$$
P_s(a*b)= P_s(a)P_s(b). 
$$ 

For a holomorphic function $f$ on the unit disk $D$, 
let $f(w)=\sum_k a_kw^k$. We have then setting 
$\pi B(s+1,m+1)= \pi\frac{\Gamma(m+1)\Gamma(s+1)}{\Gamma(s+m+2)}$, 
\begin{equation}
  \label{eq:barz}
\begin{aligned}[m]
(P_s&(\bar w^l))f(w)
=\sum_{k,m}\int_D a_k\bar v^l v^{k}\frac{1}{\pi B(s+1,m+1)}
          (w\bar v)^m(1-\bar vv)^s dv d\bar v \\ 
&= \sum_{m}\int_D a_{m+l}w^m\frac{1}{\pi B(s+1,m+1)}
   (v\bar v)^{m+l}(1-\bar vv)^s dv d\bar v \\
&= \sum_{m} a_{m+l}w^m\frac{B(s+1,m+l+1)}{B(s+1,m+1)}
 =\sum_{m} a_{m+l}w^m\frac{(m+l)\cdots (m+1)}{(s+m+l+1)\cdots(s+m+2)} 
\end{aligned}
\end{equation}
This is a bounded operator such that $(P_s(\bar w^l))w^m=0$ for 
$0\leq m\leq l-1$.  
Using \eqref{eq:barz}, we have 
\begin{equation}
  \label{eq:barzz}
  \begin{aligned}[m]
(P_s(&\bar w\!*\!w)f)(w)= (P_s(\bar w)P_s(w))f(w) \\
=& \sum_{k,m}\int_D a_k\bar v v^{k+1}\frac{1}{\pi B(s+1,m+1)}
   (w\bar v)^m(1-\bar vv)^s dv d\bar v \\
=& \sum_{k\geq 0}\int_D a_k w^k(\bar v v)^{k+1}(1-\bar vv)^s 
       \frac{1}{\pi B(s+1,k+1)}
=\sum_k a_k\frac{B(s+1,k+2)}{B(s+1,k+1)}w^k\\
=& \sum_k a_k \frac{k+1}{s+k+2} w^k 
  \end{aligned}
\end{equation}
Hence, $P_s(\bar w*w)$ is a bounded diagonal operator, which is
invertible within bounded operators. 
Note also that 
$$
P_s(\bar w*w)= P_s(\bar w\cdot w)\not= P_s(w\! *\!\bar w),\quad 
P_s((\bar w\cdot w)^k)\not=P_s((\bar w\cdot w)_*^k).
$$
Denote by $(\bar w*w)_*^{-1}$ the inverse of $\bar w*w$. Then, 
$w^{\bullet}=(\bar w*w)_*^{-1}*\bar w$ 
(resp. $\bar w^{\circ}=w*(\bar w*w)_*^{-1}$) is a left
(resp. right) inverse of $w$ (resp. $\bar w$): 
\begin{equation}
 \label{eq:hinv} 
 \begin{aligned}
((\bar w\!*\!w)_*^{-1}*\bar w)*w &= 1,
   \quad w*((\bar w\!*\!w)_*^{-1}*\bar w)=1-\varpi_0,\\    
 \bar w*(w*(\bar w\!*\!w)_*^{-1})&= 1,\quad 
(w*(\bar w\!*\!w)_*^{-1})*\bar w= 1- \varpi_0
  \end{aligned}
\end{equation}
where $\varpi_0$ is the element corresponding to the projection 
operator given by  $P_s(\varpi_0)(f)= f(0)$. 

\medskip
By \eqref{eq:barzz} we have that 
$$
(P_s(1-\bar w\!*\!w))^{-1}f(w)=(P_s((1-\bar w\! *\! w)_*^{-1})f)(w) 
    = \sum_k a_k w^k \frac{s+k+2}{s+1}
$$
This is an unbounded operator. Using this, we have  
\begin{equation}
  \label{eq:conj}
P_s((1- \bar w\! *\! w)_*^{-1}*\bar w)f(w)= \frac{1}{s+1}f'(w),\quad s\geq 0.  
\end{equation}
Note that 
$P_s((1- \bar w\! *\! w)_*^{-1}*\bar w)
     = \frac{s}{s+1}P_s(\frac{\bar w}{1-w\bar w})$.  

By \eqref{eq:hinv}, we see that 
$w*(\bar w\! *\! w)^{-1}*(1-\bar w\! *\! w)= w*(\bar w\! *\! w)^{-1}-w$ is a right 
inverse of $(1- \bar w\! *\! w)_*^{-1}*\bar w$. Indeed, we see that 
$$
P_s(w*(\bar w\! *\! w)^{-1}-w)= (s+1)\int_0^w dw.
$$

For each $s>0$, let $\Bbb D_{(s)}$ be the algebra of operators 
generated by  
$w,\bar w,$ $(\bar w*w)^{-1},$ $(1-\bar w*w)^{-1}$, $(1-w*\bar w)^{-1}$, $\h$.   

\begin{prop}\label{lem3.}
$w^*=(1-\bar w\! *\! w)_*^{-1}*\bar w$ is a canonical conjugate of $w$
i.e. $[w^{*},w]{=}\frac{1}{s{+}1}$, and 
$$
[w,\bar w]= -\frac{1}{s+1}(1-w\! *\!\bar w)*(1-\bar w\! *\! w).
$$
Hence the Berezin algebra $\Bbb D_{(s)}$ 
may be viewed as a deformation quantization of the K{\"a}hler
structure defined on the unit disk $D$ by the Poisson bracket 
 $\{w,\bar w\}{=}-\frac{1}{s+1}(1-w\bar w)^2$.
\end{prop}

\par\noindent
{\bf Proof\,\,}
Note that $-\frac{1}{s+1}= [w,(1-\bar w\! *\! w)^{-1}*\bar w]$. Hence we have 
$$
-\frac{1}{s+1}=(1-\bar w\! *\! w)_*^{-1}*[w,\bar w]*w*(1-\bar w\! *\! w)_*^{-1}*\bar w 
+ (1-\bar w\! *\! w)^{-1}*[w,\bar w]
$$
By the bumping lemma, the right hand side is  
$$
(1-\bar w\! *\! w)_*^{-1}*[w,\bar w]*((1-w\! *\!\bar w)_*^{-1}*w\! *\!\bar w +1)
= (1-\bar w\! *\! w)_*^{-1}*[w,\bar w]*(1-w\! *\!\bar w)_*^{-1}
$$
This give the result. \hfill $\Box$

\medskip
Note also that $\varpi_0$ can be obtained by using \eqref{eq:Berezin} as  
$$
\varpi_0 = \lim_{t\to\infty}e_*^{-tw*\frac{\bar w}{1-\bar ww}}
= \lim_{t\to\infty}e^{-\frac{t}{s}w\partial_w}. 
$$

\bigskip
Note here that the algebra above is given directly as
operators. Indeed, this is a deformation quantization of Poincar{\'e} disk 
in the notion that mentioned in \S\,\ref{prelim2}. 
However, in the previous sections, the concrete calculation is done by 
using $*_{_K}$-product by using an expression parameter $K$. 

\subsection{Embeddings  into the extended Weyl algebra}
On the other hand, one can easily make a deformation quantization 
of the K{\"a}hler structure on the Poincar{\'e} disk $D$ in 
${\mathcal E}_{2+}({\mathbb C}^3$. Let $\nu, x, y$ be a generator of
${\mathcal H}_2[\nu]$. We set $z{=}x{+}iy$, $\bar z{=}x{-}iy$. Using the
$*$-exponential function $e_*^{t\overline{z}{*}z}$ we see that 
$$
\sqrt[*]{\bar{z}{*}z}^{-1}, \quad \sqrt[*]{1{+}\bar{z}{*}z}^{-1} 
$$
are welldefined in ${\mathcal E}_{2+}({\mathbb C}^3)$. We set as follows:
\begin{equation} 
w{=}z{*}\sqrt[*]{\bar{z}{*}z}^{-1},\quad \overline{w}{=}\sqrt[*]{\bar{z}{*}z}^{-1}{*}\bar{z}.
\end{equation}
It is not hard to obtain 
$$
[w,\overline{w}]{=}-2\nu{*}(1{-}w{*}\overline{w}){*}(1{-}\overline{w}{*}w).
$$
Hence, we obtain a deformation quantization 
of the K{\"a}hler structure on the Poincar{\'e} disk $D$ as a
$\nu$-regulated algebra. We denote by $\Bbb D_{(\nu)}$ the 
obtained $\nu$-regulated algebra. 
The Berezin algebra  $\Bbb D_{(s)}$ gives an operator representation 
of $\Bbb D_{(\nu)}$, but the positivity of $\nu$ is implicitly assumed
in advance in the formula of beta functions.  

\bigskip
Now consider this in ${\mathcal E}_{2+}({\mathbb C}^4)$, and let  
$x_0, y_0, x_1, y_1$ be a generator system.
$$
\mu=e_*^{-2y_0},\quad \tau=e_*^{y_0}{*}x_0, \quad
\xi=e_*^{y_0}{*}x_1,\quad \eta=e_*^{y_0}{*}y_1.  
$$
Hence, its Heisenberg vacuum representation obtained in
\S\,\ref{embedded} gives an embedding of ${\mathcal H}_{2}[\mu]$ into 
 ${\mathcal E}_{2+}({\mathbb C}^4)$ such that $\mu^{-1}\to 1{+}2\h s{>}0$
together with other commutation relations. 

\medskip
As above, we set $\zeta{=}\xi{+}i\eta$,
$\overline{\zeta}{=}\xi{-}i\eta$. Then, the same calculation as above 
gives gives exactly the Berezin algebra  $\Bbb D_{(2s)}$.

\end{document}